\documentclass[aps,prd,
preprintnumbers,nofootinbib,superscriptaddress]{revtex4}
\usepackage[dvips]{graphicx}
\usepackage{bm,latexsym,amsmath,amssymb,amsfonts,color}

\DeclareMathOperator{\sech}{sech}

\begin{document}

\title{Gravitating superconducting solitons in the (3+1)-dimensional
Einstein gauged non-linear $\sigma $-model}
\author{Fabrizio Canfora}
\email{canfora@cecs.cl}
\affiliation{Centro de Estudios Cient\'{\i}ficos (CECS), Casilla 1469, Valdivia, Chile}
\author{Alex Giacomini}
\email{alexgiacomini@uach.cl}
\affiliation{Instituto de Ciencias F\'isicas y Matem\'aticas, Universidad Austral de
Chile, Casilla 567, Valdivia, Chile}
\author{Marcela Lagos}
\email{marcelagosf@gmail.com}
\affiliation{Instituto de Ciencias F\'isicas y Matem\'aticas, Universidad Austral de
Chile, Casilla 567, Valdivia, Chile}
\author{Seung Hun Oh}
\email{shoh.physics@gmail.com}
\affiliation{Seoul National University of Science and Technology, Seoul 01811, Korea}
\author{Aldo Vera}
\email{aldo.vera@uach.cl}
\affiliation{Centro de Estudios Cient\'{\i}ficos (CECS), Casilla 1469, Valdivia, Chile}
\affiliation{Instituto de Ciencias F\'isicas y Matem\'aticas, Universidad Austral de
Chile, Casilla 567, Valdivia, Chile}

\begin{abstract}
In this paper, we construct the first analytic examples of $(3+1)$%
-dimensional self-gravitating regular cosmic tube solutions which are
superconducting, free of curvature singularities and with non-trivial
topological charge in the Einstein-$SU(2)$ non-linear $\sigma$-model. These
gravitating topological solitons at a large distance from the axis look like
a (boosted) cosmic string with an angular defect given by the parameters of
the theory, and near the axis, the parameters of the solutions can be chosen
so that the metric is singularity free and without angular defect. The
curvature is concentrated on a tube around the axis. These solutions are
similar to the Cohen-Kaplan global string but regular everywhere, and the
non-linear $\sigma $-model regularizes the gravitating global string in a
similar way as a non-Abelian field regularizes the Dirac monopole. Also,
these solutions can be promoted to those of the fully coupled
Einstein-Maxwell non-linear $\sigma $-model in which the non-linear $\sigma $%
-model is minimally coupled both to the $U(1)$ gauge field and to General
Relativity. The analysis shows that these solutions behave as
superconductors as they carry a persistent current even when the $U(1)$
field vanishes. Such persistent current cannot be continuously deformed to
zero as it is tied to the topological charge of the solutions themselves.
The peculiar features of the gravitational lensing of these gravitating
solitons are shortly discussed.
\end{abstract}

\maketitle

\newpage

\section{Introduction}

Topological defects are formed in phase transitions when a system goes from
a state of higher symmetry to a state of lower symmetry. They can be
classified as local or global depending on the fact if a local or global
symmetry is broken. Topological defects occur in very different areas of
physics like, for example, condensed matter physics, high energy physics,
and cosmology. In condensed matter physics perhaps the most famous and
intuitive example is the formation of domain walls in ferromagnetic
materials, which separate domains with different magnetization. Another
famous example is the formation of vortex lines in superfluid helium and
line defects (dislocations) in crystals. A topological defect can sometimes
be completely regular in which case it is known as a topological soliton.
They play a fundamental role in quantum field theory, nuclear physics, and
high energy physics \cite{1}, \cite{2}.

The first example of stable topological soliton in three space dimensions
was proposed by Skyrme and it is known as Skyrmion \cite{skyr}. It has the
remarkable property of possessing Fermionic excitations despite the fact
that the dynamical field is an $SU(2)$-valued scalar field. At leading order
in the 't Hooft large \textbf{N} expansion \cite{5}, \cite{6}, \cite{7}, the
Skyrme model represents a (phenomenologically successful) low energy
description of Quantum Chromodynamics (QCD).

The Skyrme model arose from a clever modification of the non-linear $\sigma $%
-model (NLSM henceforth), which is the low energy description of the
dynamics of Pions (for nice reviews see e.g. \cite{3}, \cite{4}). The Skyrme
term was added to the NLSM in order to avoid Derrick's no-go scaling
argument \cite{8} preventing the existence of static soliton solutions of
finite energy. On the other hand, it should be kept in mind that the elegant
arguments in \cite{9} (see also \cite{11}, \cite{12}, \cite{13}, \cite{14},
and references therein) to show that Skyrmions represent Fermions (at least
semi-classically) are only based on the existence of stable solitons with
non-trivial third homotopy class while they do not use directly the Skyrme
term in itself. Thus, it is extremely interesting to search for alternative
ways to avoid Derrick's scaling argument in order to achieve
\textquotedblleft \textit{Fermions out of Bosons}" with the simplest
possible ingredients. The main approaches to avoid the no-go scaling
argument in \cite{8} are, first of all, to minimally couple the NLSM to
gravity and/or to Maxwell theory. Indeed, using the techniques developed in 
\cite{canfora2}, \cite{56}, \cite{56b}, \cite{58}, \cite{58b}, \cite{ACZ}, 
\cite{Ayon2}, \cite{CanTalSk1}, \cite{canfora10}, \cite{Giacomini:2017xno}, 
\cite{ACLV}, \cite{Fab1}, \cite{gaugsk}, \cite{Canfora:2018clt}, \cite%
{lastEPJC}, \cite{crystal1}, \cite{crystal2}, exact self-gravitating NLSM
solutions with non-trivial topological charge have been found in \cite{CDP}, 
\cite{CDGP} and \cite{GO}. Secondly, it is very helpful to construct
time-dependent ansatz for the $SU(2)$-valued matter field with the property
that the energy-momentum tensor is time-independent (this idea is the $SU(2)$
generalization of the Bosons star ansatz for a $U(1)$-charged scalar field:
see \cite{62}, \cite{63} and references therein). Such a generalization has
been achieved in \cite{ACZ}, \cite{Ayon2}, \cite{Fab1}, \cite{crystal1} and 
\cite{crystal2}.

Further relevant topological solitons which play an important role in high
energy as well as condensed matter physics, are the Abrikosov-Nielsen-Olesen
vortex line \cite{abrikosov}, \cite{Nie-Ole} and the 't Hooft-Polyakov
monopole in the $SU(2)$ Yang-Mills-Higgs system \cite{thooft}, \cite%
{polyakov}. It is worth to mention that the latter at large distances looks
like a Dirac monopole, however, the Dirac monopole, which describes a
point-like magnetic charge, is singular at the origin whereas the 't
Hooft-Polyakov monopole is regular at the origin.\newline

The formation of topological defects are very important in grand unified
theories as the actual symmetry group of the standard model is supposed to
be a result of a series of spontaneous symmetry breaking of a larger
symmetry group. This means that topological defects play a fundamental role
from microscopic scale to extremely large scale namely in cosmology due to
the fact that the universe in its evolution expanded and cooled down, and
therefore went through several phase transitions were topological defects
have formed. As in cosmology, the most relevant interaction is gravity, the
topological defects must be studied in the context of field theories coupled
to gravity.

In cosmology the topological defects which attracted the most attention of
the scientific community are the cosmic strings. The simplest exact cosmic
string solution is given by an energy-momentum tensor concentrated in a line 
\cite{vilenkin1} (for example the $z$ axis) 
\begin{equation}
T_{\alpha \beta }\ =\ \mu \delta (x)\delta (y)\mathrm{diag}(1,0,0,1)\ .
\end{equation}%
The exact solution of the Einstein field equations associated with this
energy-momentum tensor, written in cylindrical coordinates, is locally but
not globally flat 
\begin{equation}
ds^{2}=-dt^{2}+dr^{2}+r^{2}d\theta ^{2}+dz^{2}\ ,
\end{equation}%
as the range of the angular coordinate $\theta $ is not, as usual, $2\pi $
but 
\begin{equation}
0\leq \theta \leq 2\pi (1-4G\mu )\ ,
\end{equation}%
where $G$ and $\mu $ are the gravitational constant and the mass of the
cosmic string per unit length, respectively. Therefore, this locally flat
space-time has a conical defect, an angular deficit $\Delta =8G\mu $ and has
a curvature singularity on the $z$ axis. A possible way to smooth out the
singularity is to smear the energy-momentum tensor on a cylinder of finite
radius $\delta $ and it is possible to find exact solutions \cite{gott}, 
\cite{hiscock}, \cite{linet}. The principal problem of this procedure is
that there is a sharp boundary whose radius is arbitrary and moreover the
energy-momentum tensor is not derived from some fundamental action principle.

In order to find a cosmic string solution from a fundamental action
principle usually the Einstein-Yang-Mills-Higgs action is used, but
unfortunately no exact solutions are known. An explicit gravitating global
cosmic string metric (where the fundamental field is a Goldstone boson
instead of the Higgs and Yang-Mills fields) was found by Cohen and Kaplan,
and has the peculiarity that the matter distribution has no sharp boundary
and the angular defect varies with the distance from the symmetry axis \cite%
{cohen-kaplan}. This metric has a curvature singularity at a finite distance
from the axis. It was shown that non-singular gravitating global strings can
exist if there is an explicit time dependence in the metric \cite{gregory}, 
\cite{draper}. For a nice explanation of the properties of the Cohen-Kaplan
global string see e.g. \cite{bookVS} pages 196-198.

The existence of cosmic strings has many important cosmological and
astrophysical implications. For example, cosmic strings have been proposed
to have a role in the galaxy formation as a source of density perturbations 
\cite{kibble}. Cosmic strings have also observable effects through
gravitational lensing being the most known effect the formation of double
images \cite{vilenkin1}, \cite{vilenkin3}, \cite{gott2}. It is worth to
point out that the space-time generated by a thin string (with a Dirac delta
matter source) does not exert force on a test particle being locally flat,
but the existence of a conical defect still generates double images. Perhaps
one of the most fascinating aspects of cosmic strings is that, under certain
conditions, they become superconducting as it was shown in the pioneering
article of Witten \cite{wittenstrings}. This can have observable effects as
such superconducting strings would act as sources of synchrotron radiation
or high energy cosmic rays \cite{vilenkin2}. Moreover, it has also been
proposed that superconducting strings moving in a magnetized plasma can be a
mechanism for the production of gamma-ray bursts \cite{BHV}.\newline

Due to the many important cosmological and astrophysical implications of
cosmic strings, it is of great interest to find analytic non-singular
solutions that can be derived from some fundamental action principle which
leaves no arbitrariness in the choice of fields and their potentials.
Indeed, as it has been explained before, in the case of the
Einstein-Yang-Mills-Higgs action no such exact solutions are known. The
known exact solutions are the thin string with Dirac delta matter source and
the global string, both of them possess curvature singularities which would
go against the cosmic censorship conjecture. An important point is how to
choose a fundamental matter field. Here we will consider the (gauged) NLSM
as it is an effective low energy description of the dynamics of Pions (as
well as of their electromagnetic properties).

In this paper, we construct the first examples of analytic and singularity
free cosmic tube solutions for the self-gravitating $SU(2)$-NLSM. At large
distance the metric behave in a similar way to the one of a cosmic string
boosted in the axis and has an angular defect related to the parameter of
the theory, however near the axis the parameter of the solution can be
chosen in such way that the solution is free of singularities and without
angular defect. This means that these solutions are free of singularities
everywhere and the angular defect depends on the distance from the axis. It
is also worth pointing out that the matter field does not have a sharp
boundary and the curvature reaches its maximum on a tube around the axis
rather than on the axis itself. All these features make the new solutions
similar to a global string with the big difference that they are regular
instead of having a singularity at finite distance from the axis. The cosmic
tubes found here are related to global strings in the same way as
non-Abelian monopole are related to Dirac monopoles. In other words, the
NLSM regularizes the global string keeping a similar behavior at large
distances. Another relevant feature of the solutions presented here is that
these possess non-trivial topological charge (the third homotopy class). It
is important to mention that in \cite{Jackson:1988xk}, \cite{Jackson:1988ku}
and \cite{Nitta:2007zv}, string solutions in the Skyrme model with mass term
have been constructed. These configurations were obtained using a static
ansatz that facilitates the resolution of the field equations numerically,
but that leads to the strings possessing a zero topological density. Those
strings, therefore, are classically unstable and are expected to decay into
Pions. On the other hand, the temporal dependency in the $U$ field that we
introduce, as we will see before, allows us to construct analytical and
topologically non-trivial solutions by circumventing Derrick's theorem (as
we have previously pointed out) and without having to include the Skyrme
term or additional potentials. Our solutions, having a topological charge,
can not decay in the configurations found in \cite{Jackson:1988xk}, \cite%
{Jackson:1988ku} and \cite{Nitta:2007zv}. Furthermore, these tubes can be
promoted to full solutions of the Einstein-Maxwell-NLSM in which the NLSM
field is minimally coupled both to the $U(1)$ field as well as to gravity.
These gauged solutions carry a persistent current even when the $U(1)$ gauge
field is zero and therefore becomes superconducting in the sense of \cite%
{wittenstrings}. In particular, the superconducting currents are tied to the
topological charge so that they cannot be deformed continuously to zero
(that is why they are persistent). It is worth to emphasize that the
gravitating solitons constructed in the present paper only involve degrees
of freedom arising from low energy QCD minimally coupled with General
Relativity (GR) without the need of additional potentials.

\subsection{Comparison with Existing Literature}

Here it is interesting to compare our novel results with some of the results
already existing in the literature. In particular, we will refer to the nice
references \cite{Jackson:1988xk}, \cite{Jackson:1988ku}, \cite{Nitta:2007zv}%
, \cite{Volkov1}, \cite{Volkov2}, \cite{Comtet:1987wi}, \cite{Moss:1987ka}, 
\cite{Babul:1988qt}, \cite{Zhang:1997is}, \cite{Balachandran:2001qn}, \cite%
{Carter:2002te}. Here below we list these four important new features of our
solutions (for the sake of clarity, we will call these novel features 
\textbf{A}, \textbf{B}, \textbf{C} and \textbf{D}).

\textbf{A)} First and foremost, our regular and topologically non-trivial
superconducting tubes are also gravitating. Namely, we couple them to the
Einstein equations since one of our main goals is to determine the peculiar
features of the gravitational field generated by superconducting tubes. The
reason is that, as it was already clarified in Witten's original reference,
the spectacular observational effects of superconducting tubes appear in
situations in which the gravitational field should not be neglected.
Moreover (for the reasons explained here below in point B) we are willing to
construct such gravitating superconducting tubes analytically rather than
numerically. On the other hand, many of the relevant examples of
superconductive tubes available in the literature are numerical and found in
flat space-times (such as the examples in the very nice references \cite%
{Jackson:1988xk}, \cite{Jackson:1988ku}, \cite{Nitta:2007zv}, \cite{Volkov1}%
, \cite{Volkov2}, \cite{Balachandran:2001qn}).

\textbf{B)} The second main novel feature is that our regular gravitating
solitons in a sector with non-vanishing topological charge are completely
analytic: to the best of our knowledge, all the most relevant examples
available in the literature are numerical\footnote{%
Analytic non-topological cosmic strings have been studied in other different
non-linear models, see for example \cite{Comtet:1987wi}, \cite{Moss:1987ka}
and \cite{Babul:1988qt}.}. An obvious question is: why should one insist so
much on finding analytic solutions if these equations can be solved
numerically? Indeed, very powerful numerical techniques are available in the
literature to analyze gravitating superconducting tubes (as, for instance,
in \cite{Jackson:1988xk}, \cite{Jackson:1988ku}, \cite{Nitta:2007zv}, \cite%
{Volkov1}, \cite{Volkov2}, \cite{Comtet:1987wi}, \cite{Moss:1987ka}, \cite%
{Babul:1988qt}, \cite{Zhang:1997is}, \cite{Balachandran:2001qn}, \cite%
{Carter:2002te}). There are many sound reasons to strive for analytic
solutions even when numerical techniques are available (especially when, as
in the present paper, one is working in $(3+1) $-dimensions). Here below we
will list two of them which are quite relevant for the present analysis.
First of all, as general motivation, it could be enough to remind all the
fundamental concepts that we have understood thanks to the availability of
the Schwarzschild and Kerr solutions in GR and of the non-Abelian monopoles
and instantons in Yang-Mills-Higgs theory. Much of what we now know about
black hole physics in GR and instantons and monopoles in gauge theories
arose from a careful study of the available analytic solutions. Hence, a
systematic tool to construct analytic gravitating superconducting tubes can
greatly enlarge our understanding of these very important topological
defects. Secondly, the analytic knowledge of these configurations greatly
simplifies the gravitational lensing analysis (indeed, it is worth
emphasizing that gravitational lensing is the main phenomenological tool
with which one can discover such objects).

\textbf{C)} The third novel feature of our gravitating superconducting tubes
is that they are constructed entirely in the low energy limit of QCD
minimally coupled both to GR and to Maxwell theory\footnote{%
The existence of superconducting strings in the low energy limit of QCD has
been discussed in the nice works \cite{Zhang:1997is}, \cite%
{Balachandran:2001qn} and \cite{Carter:2002te}}. Indeed, at leading order in
the large \textbf{N} limit, the $SU(2)$ NLSM is the low energy limit of QCD
describing the dynamics of Pions (and also of Baryons when the topological
charge is non-vanishing as in our manuscript). Thus, in particular, our
solutions are physically very different from the ones in the two nice
references \cite{Volkov1}, \cite{Volkov2} (which are closely related to the
electroweak sector of the standard model). Our solutions instead can be very
relevant in situations in which Hadrons interact strongly both
electromagnetically and gravitationally. Needless to say, there are many
relevant situations from the astrophysical and cosmological point of view
where both the electromagnetic and the gravitational interactions with
Hadrons cannot be neglected (as it happens, for instance, in the physics of
neutron stars). Hence, our solutions are not only of academic interest but
they can play an important role in phenomenologically relevant situations.

\textbf{D)} The fourth novel feature of our gravitating superconducting
tubes is that the persistent character of the superconductive current is
extremely transparent. In particular, we have taken great care to build an
ansatz with the following properties. Firstly, we want a consistent ansatz
with a non-vanishing topological charge (which, in our case, is interpreted
as the Baryonic charge of the configuration) because configurations with
non-vanishing topological charge cannot decay into the trivial
configurations (neither, in fact, into configurations with different
topological charges). Secondly, we want a consistent ansatz in which the
electromagnetic current is protected by topology (for instance, the
numerical solutions constructed in the nice references by Nitta-Shiki and
Jackson have vanishing topological charge \cite{Jackson:1988xk}, \cite%
{Jackson:1988ku}, \cite{Nitta:2007zv}). Thirdly, we need a time-dependent
ansatz (in order to avoid the Derrick scaling argument: see \cite{Rosen:1968}%
, \cite{Friedberg:1979} and \cite{Coleman:1985}.) which is, at the same
time, compatible with a stationary metric and carries a non-trivial
topological charge.

In other words, the electromagnetic current associated with our
configurations is proportional to the topological density of the
configurations themselves. The physical consequence of this fact is that our
current cannot be deformed continuously to the trivial vanishing current as
this would imply also a jump in the topological charge. However, there is no
continuous deformation that can change the topological charge. It is worth
to emphasize that the energy scale of GUT symmetry breaking is $10^{15}$ GeV
while the non-linear $\sigma$-model considered in the present manuscript is
motivated as a low-energy effective theory of QCD, is relevant at much lower
energy scales. On the other hand, the superconducting nature of the
gravitating solitons to be constructed here below may strongly affect the
scattering of electromagnetic waves leading to potentially visible effects.
The reason is that gravitational lensing does not take into account the very
strong interactions of the electromagnetic field with the matter field.
Unfortunately, we cannot be precise yet since the problem of the scattering
of electromagnetic waves in these gravitating solitons background must be
solved numerically. We hope to come back to this interesting point in a
future publication. Therefore, our superconducting currents are entirely
protected by topology and they cannot decay; these are genuine persistent
(and, therefore, superconductive) currents. Moreover, in the original
Witten's reference, superconducting currents can only appear when suitable
inequalities on the parameters of the Higgs potential are satisfied. In our
case instead we do not need any potential at all; it is just the topology of
the low energy limit of QCD which keeps the currents alive forever without
any extra ingredient.

The structure of the paper is the following: in Sections II and III, we
present the model, give our ansatz and the corresponding field equations. In
Section IV we construct the exact regular cosmic tube solutions of the
Einstein $SU(2)$-NLSM and show that they possess non-trivial topological charge. Also
we study the geodesic equations and their main physical properties. In Section V
it will be shown how to promote the found configurations to be solutions of the
gauged Einstein NLSM system and how these are superconducting
configurations. In Section VI we discuss the flat limit of our configurations. In the last Section, our conclusions are detailed.

\section{The Einstein $SU(2)$-NLSM}

The Einstein-NLSM theory is described by the action 
\begin{equation}
I[g,U]=\int d^{4}x\sqrt{-g}\left( \frac{\mathcal{R}}{2\kappa }+\frac{K}{4}%
\mathrm{Tr}[L^{\mu }L_{\mu }]\right) ,  \label{action1}
\end{equation}%
where $\mathcal{R}$ is the Ricci scalar and $L_{\mu }$ are the Maurer-Cartan
form components $L_{\mu }=U^{-1}\partial _{\mu }U$ for $U(x) \in SU(2)$.
Here $\kappa $ is the gravitational constant and the positive coupling $K$
is fixed by experimental data. In our convention $c=\hbar =1$ and Greek
indices run over the four dimensional space-time with mostly plus signature.
In this paper, we follow a standard convention that the Riemann curvature
tensor, the Ricci tensor, and the Ricci scalar are given by 
\begin{gather*}
R^{\alpha}_{\phantom{\alpha} \beta \mu \nu} = \Gamma^{\alpha}_{%
\phantom{\alpha} \beta \nu , \mu} -\Gamma^{\alpha}_{\phantom{\alpha} \beta
\mu, \nu} +\Gamma^{\alpha}_{\phantom{\alpha} \rho \mu} \Gamma^{\rho}_{%
\phantom{\rho} \beta \nu} -\Gamma^{\alpha}_{\phantom{\alpha} \rho \nu}
\Gamma^{\rho}_{\phantom{\rho} \beta \mu} \ , \\
R_{\mu \nu} = R^{\alpha}_{\phantom{\alpha} \mu \alpha \nu} \ , \qquad 
\mathcal{R} = g^{\mu \nu} R_{\mu \nu} \ ,
\end{gather*}
respectively.

In order to produce a correct physical interpretation of the topological
solitons constructed here and compare with the cosmic string solutions
already existing in the literature, in this work we will refer to
configurations with vanishing cosmological constant, however the techniques
used here are also effective in presence of a cosmological constant. We hope
to come back on this interesting issue in a future publication.

The complete Einstein-NLSM equations read 
\begin{equation}
\nabla ^{\mu }L_{\mu }=0\ ,\qquad G_{\mu \nu }=\kappa T_{\mu \nu }\ ,
\label{skyrme-einstein}
\end{equation}%
where $G_{\mu \nu }$ is the Einstein tensor and $T_{\mu \nu }$ the
energy-momentum tensor of the NLSM given by 
\begin{equation}  \label{TmunuNLSM}
T_{\mu \nu }=-\frac{K}{2}\mathrm{Tr}\left[ L_{\mu }L_{\nu }-\frac{1}{2}%
g_{\mu \nu }L^{\alpha }L_{\alpha }\right] .
\end{equation}%
The winding number of the configurations reads 
\begin{equation}
w_{\text{B}}=\frac{1}{24\pi ^{2}}\int \rho _{\text{B}}\ ,\quad \rho _{\text{B%
}}=\text{Tr}[\epsilon ^{ijk}L_{i}L_{j}L_{k}]\ .  \label{winding}
\end{equation}%
When the topological density $\rho _{\text{B}}$ is integrated on a
space-like surface $w_{\text{B}}$ represents the Baryon number. We will only
consider configurations in which $\rho _{\text{B}}\neq 0$.

To construct analytical solutions in this theory we will use the generalized
hedgehog ansatz (see \cite{canfora2}, \cite{56}, \cite{56b}, \cite{58}, \cite%
{58b}, \cite{ACZ}, \cite{Ayon2}, \cite{CanTalSk1}, \cite{canfora10}, \cite%
{Giacomini:2017xno}, \cite{ACLV}, \cite{Fab1}, \cite{gaugsk}, \cite%
{Canfora:2018clt}, \cite{lastEPJC}, \cite{crystal1}, \cite{crystal2}), which
is defined as 
\begin{gather}
U^{\pm 1}(x^{\mu })=\cos \left( \alpha \right) \mathbf{1}_{2}\pm \sin \left(
\alpha \right) n^{i}t_{i}\ ,\ \ \ \ n^{i}n_{i}=1\ ,  \label{hedgehog} \\
n^{1}=\sin \Theta \cos \Phi \ ,\ \ \ n^{2}=\sin \Theta \sin \Phi \ ,\ \ \
n^{3}=\cos \Theta \ ,  \notag
\end{gather}%
where $t_i \equiv i \sigma_i$, with $\sigma_i$ are the Pauli matrices. It
should be noticed that the Maurer-Cartan forms are adapted for the process
of solving the field equations, since they satisfy the $SU(2)$ algebra. The
choice of the ansatz in Eq. \eqref{hedgehog} is adapted to the commutation
relations of the Maurer-Cartan form in such a way to simplify as much as
possible the corresponding Einstein equations (as it will be discussed in a
moment). A necessary (but, in general, not sufficient) condition in order to
have non-vanishing topological charge is%
\begin{equation}
d\alpha \wedge d\Theta \wedge d\Phi \neq 0 \ , \qquad \Longleftrightarrow
\qquad \epsilon^{ijk} (\partial_{i} \alpha) (\partial_{j} \Theta)
(\partial_{k} \Phi) \neq 0 \ ,  \label{cond01}
\end{equation}%
where $\alpha $, $\Theta $\ and $\Phi $\ are the three scalar degrees of
freedom appearing in the standard parametrization of the $SU(2)$-valued
scalar field defined in Eq. (\ref{hedgehog}).

\section{Ansatz and Field Equations}

The ansatz in Eq. \eqref{hedgehog} is defined over the Weyl-Lewis-Papapetrou
metric 
\begin{equation}
ds^{2}=-\frac{B_{0}^{2}\omega _{s}}{e^{f_{0}}}\Big(\frac{2e^{f_{0}}}{B_{0}}%
-\omega _{s}G\Big)dt^{2}-\frac{2B_{0}^{2}}{e^{f_{0}}}\Big(\frac{e^{f_{0}}}{%
B_{0}}-\omega _{s}G\Big)dtdz +\frac{B_{0}^{2}}{e^{f_{0}}}Gdz ^{2}+e^{-2R}%
\big(dX^{2}+d\theta ^{2}\big)\ ,  \label{WLP}
\end{equation}%
with $G=G(X,\theta )$, $R=R(X)$, while $B_{0}$, $\omega _{s}$, $f_{0}$ are
arbitrary constants. Since the metric determinant of the section spanned by $%
t$ and $z $ is negative definite (and $e^{-2R}$ is definite positive as we
will show in the next sections), 
\begin{equation}
\begin{vmatrix}
g_{tt} & g_{tz } \\ 
g_{z t} & g_{z z }%
\end{vmatrix}%
=-B_{0}^{2}<0 \ ,
\end{equation}%
the space-time with this metric is always Lorentzian regardless of the
metric components in the $t-z $. As it will be discussed in the next
sections, in order to have an integer value of the topological charge, one
must allow the whole range of real number as the domain of the coordinate $X$%
: 
\begin{equation}
-\infty <X<\infty \ ,
\end{equation}%
and $\theta $ is an angular coordinate with range 
\begin{equation*}
\theta \in \lbrack 0,2\pi ] \ .
\end{equation*}
According to \cite{crystal1}, \cite{crystal2} we will consider a matter
field in the form 
\begin{equation}
\alpha =\alpha (X)\ ,\qquad \Theta =q\theta \ ,\qquad \Phi =\omega _{s}t+z \
.  \label{ansatz}
\end{equation}%
In Appendix \ref{App1} we have given a detailed explanation of how the above
ansatz can be constructed in a systematic way and why it is so effective in
the case of the gauged gravitating NLSM. Note that the ansatz defined here
above satisfies the necessary condition in Eq. (\ref{cond01}) in order to
possess a non-trivial topological charge. The sufficient conditions will be
discussed in the following sections.

This ansatz is very useful for (at least) three reasons. The first one is
because Eq. \eqref{ansatz} implies the relations 
\begin{equation*}
\nabla _{\mu }\Phi \nabla ^{\mu }\Phi =0\ ,\qquad \nabla _{\mu }\Theta
\nabla ^{\mu }\Phi =0\ ,
\end{equation*}%
which simplify greatly the NLSM equations. The second reason is that Eq. %
\eqref{ansatz} allows to avoid the Derrick's scale argument \cite{8} as it
is a time dependent ansatz which, however is compatible with a stationary
metric. Thirdly, the three coupled field equations for the NLSM in the
metric defined in Eq. (\ref{WLP}) reduce to the single second order ODE for $%
\alpha (X)$: 
\begin{equation}
\alpha ^{\prime \prime }-\frac{q^{2}}{2}\sin (2\alpha )\ =0\ .
\label{Eqalpha}
\end{equation}%
It is important to note that this second order equation for $\alpha (X)$ can
be reduced to the following first order equation 
\begin{equation}
(\alpha ^{\prime })^{2}-q^{2}\sin ^{2}{\alpha }\ =E_{0}\ ,  \label{Eqalpha0}
\end{equation}%
where $E_{0}$ is an integration constant. However, the compatibility with
the Einstein equation requires that $E_{0}=0$. This situation is different
from what happens in flat space-time, where
the integration constant $E_{0}$ can be non-zero, see \cite{crystal1} and \cite{crystal2}.

Quite remarkably, the Einstein equations with the energy-momentum tensor
corresponding to the NLSM configuration defined in Eq. (\ref{ansatz}) are
reduced to only two solvable equations: 
\begin{align}
R^{\prime \prime }-K\kappa q^{2}\sin ^{2}{\alpha }\ & =\ 0\ ,  \label{EqR} \\
(\partial _{X}^{2}+\partial _{\theta }^{2})G+2C_{0}\sin ^{2}(q\theta
)e^{-2R}\sin ^{2}{\alpha }\ & =\ 0\ ,  \label{EqG}
\end{align}%
where $C_{0}=K\kappa e^{f_{0}}/B_{0}^{2}$. Note that once $R$ and $\alpha $
are solved, $G$ can be obtained directly. In fact, Eq. \eqref{EqG} is
nothing but a flat linear Poisson equation in two dimensions in which the
source term is\ known explicitly (as $\alpha (X)$ and $R(X)$ have been
determined in Eqs. (\ref{Eqalpha}) and (\ref{EqR})). Hence, Eq. \eqref{EqG}
can be solved, for instance, using the method of Green's function. In the
next sections we will construct the solution using a direct method. At this
point, it is worth to emphasize that the function $G$ depends explicitly on $%
\theta $. As the coordinate $\theta $ plays the role of an angular
coordinate going around the Hadronic tube, the fact that $G$ depends on $%
\theta $ implies that the present family of gravitating solitons is not
axi-symmetric. Indeed, our space-time admits two Killing vectors $\partial_{t}
$ and $\partial_{z}$ so that there does not exist any closed curve generated
by a spacelike Killing vector. Moreover, a direct computation reveals that%
\textit{\ the Killing equation associted to the present class of metrics in
Eqs}. (\ref{WLP}) and (\ref{ansatz}) \textit{has no further solution other
than the two solutions $\partial_{t}$ and $\partial_{z}$}: this means that
(due to the presence of the function $G(X,\theta )$) the metric is \textit{%
manifestly not axi-symmetric} as, otherwise, $\partial _{\theta }$ would be a
Killing field as well.

The physical role of the function $G$ will be discussed in the analysis of
the curvature invariants and of the geodesics.

\section{Gravitating tubes}

In this Section we construct exact regular cosmic tube solutions of the
Einstein-NLSM and we study its relevant physical properties.

\subsection{Solving the system}

Eqs. \eqref{Eqalpha} and \eqref{EqR} can be solved analytically, and the
expressions for $\alpha(X)$ and $R(X)$ are given by 
\begin{equation}
\alpha (X)=2\arctan \exp \big(q X+C_{1}\big) \ ,  \label{alphasol}
\end{equation}%
\begin{equation}
R(X)=K\kappa \ln \big(\cosh (qX+C_{1})\big)+C_{2}X+C_{3} \ ,  \label{Rsol}
\end{equation}
where $C_1$, $C_2$ and $C_3$ are integration constants. Replacing the above
in Eq. \eqref{EqG} we obtain the final equation for $G$, 
\begin{align}  \label{EqGn}
\partial_X^2 G + \partial_\theta^2 G +\frac{ 2 K\kappa}{B_0^2} e^{-2(C_2
X+C_3) +f_0} \cosh^{-2(K\kappa+1)}(q X+C_1) \sin^2(q\theta) \ = \ 0 \ .
\end{align}
At this point it is important to emphasize that considering only Eqs. %
\eqref{hedgehog}, \eqref{WLP}, \eqref{ansatz}, \eqref{alphasol} and %
\eqref{Rsol} the complete Einstein-NLSM system has been reduced to a single
equation for the metric function $G$ given in Eq. \eqref{EqGn}.

The energy density (measured by a co-moving observer) of these
configurations is given by 
\begin{align}
& T_{\hat{0}\hat{0}}=K\Big[q^{2}e^{2C_{2} X +C_{3}}\sech^{2(1-K\kappa
)}(qX+C_{1})+\frac{\omega _{s}e^{f_{0}}}{B_{0}(2e^{f_{0}}-B_{0}\omega _{s}G)}%
\sin ^{2}(q\theta )\sech^{2}(qX+C_{1})\Big] \ , \\
& \text{where}\qquad T_{\hat{0}\hat{0}}=\mathbf{T}(\hat{\mathbf{e}}_{0},\hat{%
\mathbf{e}}_{0})\qquad \text{for}\qquad \hat{\mathbf{e}}_{0}=\frac{1}{\sqrt{%
-g_{00}}}\partial _{t} \ .  \notag
\end{align}%
Note that we should impose the constraint on the integration constants 
\begin{equation}
C_{2}<(1-K\kappa )\ |q| \ ,  \label{cond1}
\end{equation}%
to avoid divergence of the energy density at $X \rightarrow \pm \infty $.

\subsection{An analytical solution}

Using that $\sin ^{2}(x)=(1-\cos (2x ))/2$ , Eq. \eqref{EqG} becomes 
\begin{equation}
(\partial _{X}^{2}+\partial _{\theta }^{2})G+C_{0}e^{-2R(X)}\sin ^{2}{\alpha
(X)}-C_{0}\cos (2q\theta )\times e^{-2R(X)}\sin ^{2}{\alpha (X)}=0\ .
\label{maineqG3}
\end{equation}%
The function $G$ can be expressed by the sum of two functions $G_{1}(X)$ and 
$G_{2}(X,\theta )$ 
\begin{equation}
G(X,\theta )=G_{1}(X)+G_{2}(X,\theta )\ ,  \label{splitG}
\end{equation}%
which satisfy 
\begin{gather}
\frac{d^{2}G_{1}}{dX^{2}}=-C_{0}e^{-2R(X)}\sin ^{2}{\alpha (X)}\ ,  \notag
\label{maineqG2-1} \\
(\partial _{X}^{2}+\partial _{\theta }^{2})G_{2}=C_{0}\cos (2q\theta )\times
e^{-2R(X)}\sin ^{2}{\alpha (X)}\ .  \label{maineqG2-2}
\end{gather}%
The first equation is trivially solved by a double integral with respect to $%
r$, 
\begin{equation}
G_{1}(X)=-C_{0}\int_{-\infty }^{X}dX_{1}\int_{-\infty }^{X_{1}}dX_{2}\
e^{-2R(X_{2})}\sin ^{2}{\alpha (X_{2})}\ .  \label{finalG0}
\end{equation}%
We can solve the equation for $G_{2}$ by separation of variables, obtaining 
\begin{equation}
G_{2}(X,\theta )=y\ \psi _{1}(X)\cos (2q\theta )\ ,
\end{equation}%
for some real constant $y$, and $\psi _{1}(X)$ satisfying 
\begin{equation}
\psi _{1}^{\prime \prime }-4q^{2}\psi _{1}=\frac{C_{0}}{y}\times
e^{-2R(X)}\sin ^{2}{\alpha (X)}\ .  \label{eqtosolve}
\end{equation}
Eq. \eqref{eqtosolve} can be solve in terms of Elliptic Functions using the
method of variation of parameters. The solution of the homogeneous equation
is 
\begin{equation*}
\psi _{\text{h}}=a\psi _{\text{h}}^{(1)}+b\psi _{\text{h}%
}^{(2)}=ae^{2qX}+be^{-2qX}\ ,
\end{equation*}%
with $a$, $b$ integration constants. On the other hand, a particular
solution can be found through 
\begin{equation*}
\psi _{\text{p}}=A(X)\psi _{\text{h}}^{(1)}+B(X)\psi_{\text{h}}^{(2)}\ ,
\end{equation*}%
where 
\begin{equation*}
A(X)=\int \frac{-\psi _{\text{h}}^{(2)}P(X)}{W(\psi _{\text{h}}^{(1)},\psi_{%
\text{h}}^{(2)})}dX \ , \quad B(X)=\int \frac{\psi_{\text{h}}^{(1)}P(X)}{%
W(\psi_{\text{h}}^{(1)},\psi_{\text{h}}^{(2)})}dX \ ,
\end{equation*}%
here $W(\psi_{\text{h}}^{(1)},\psi _{\text{h}}^{(2)})=-4q$ is the Wronskian,
and 
\begin{equation*}
P(X)=\frac{C_{0}}{y}e^{-2R}\sin ^{2}{\alpha }\ .
\end{equation*}%
Therefore, the solution of Eq. \eqref{eqtosolve} is the sum of $\psi _{\text{%
h}}$ and $\psi _{\text{p}}$, namely 
\begin{align}
\psi_{1}\ =& \ ae^{2qX}+be^{-2qX} +\frac{C_{0} e^{-2(C_{1}+qX)} \cosh
^{-2K\kappa }(C_{1}+qX)}{2yK\kappa q^{2}}  \notag \\
& \times \biggl( 1+(e^{2(C_{1}+qX)}-1) {}_{2}F_{1} [1,-1-K\kappa ,1+K\kappa
,-e^{2(C_{1}+qX)}]\biggl)\ .  \label{finalG}
\end{align}

\subsection{Regularity}

\label{regularity}

Obviously, the regularity of the metric must be discussed by analyzing the
corresponding curvature invariants. Taking into account Eqs. (\ref{alphasol}%
) and (\ref{Rsol}), the Ricci scalar $S$, the Kretschmann scalar, the square
of the Ricci tensor and of the Weyl tensor for the metric in Eq. (\ref{WLP})
read 
\begin{gather*}
S=2K\kappa q^{2}e^{2(C_{2}X+C_{3})}\cosh ^{2(K\kappa -1)}(qX+C_{1})\ ,\qquad
R^{\mu \nu \rho \sigma }R_{\mu \nu \rho \sigma }=S^{2}\ ,\qquad R^{\mu \nu
}R_{\mu \nu }=\frac{1}{2}S^{2}\ , \\
C_{\alpha \beta \gamma \delta }C^{\alpha \beta \gamma \delta }=\frac{%
4K^{2}\kappa ^{2}}{3}\Big[qe^{C_{2}X+C_{3}}\cosh ^{K\kappa -1}(qX+C_{1})\Big]%
^{4}\ ,
\end{gather*}%
and are all regular everywhere (in particular, $S$ is regular at $X
\rightarrow \pm \infty $) if we impose the following condition on the
integration constants: 
\begin{equation}
|C_{2}|<(1-K\kappa )\ |q|\ .  \label{cond2}
\end{equation}%
One may wonder whether the above condition in Eq. (\ref{cond2}) is enough to
ensure the regularity of the metric. Indeed, it is possible to compute
explicitly the main fourteen curvature invariants that are usually
considered in the literature to analyze, in four dimensions, the issue of
regularity (see \cite{Sneddon}, \cite{Harvey}). Such invariants are 
\begin{align*}
I_{1}& =S=R_{\mu }{}^{\mu }=2e^{2R}R^{\prime \prime }\ ,\quad I_{2}=R_{\mu
}{}^{\nu }R_{\nu }{}^{\mu }=\frac{1}{2}I_{1}^{2}\ ,\quad I_{3}=R_{\mu
}{}^{\nu }R_{\rho }{}^{\mu }R_{\nu }{}^{\rho }=\frac{1}{4}I_{1}^{3}\ , \\
I_{4}& =R_{\mu }{}^{\nu }R_{\rho }{}^{\mu }R_{\sigma }{}^{\rho }R_{\nu
}{}^{\sigma }=\frac{1}{8}I_{1}^{4},\quad J_{1}=A_{\mu \nu \rho \sigma
}g^{\mu \rho }g^{\nu \sigma }=\frac{1}{3}I_{1}^{2}\ ,\quad J_{2}=B_{\mu \nu
\rho \sigma }g^{\mu \rho }g^{\nu \sigma }=\frac{1}{18}I_{1}^{3}\ , \\
J_{3}& =E_{\mu \nu \rho \sigma }g^{\mu \rho }g^{\nu \sigma }=0\ ,\quad
J_{4}=F_{\mu \nu \rho \sigma }g^{\mu \rho }g^{\nu \sigma }=0\ ,\quad
K_{1}=C_{\mu \nu \rho \sigma }R^{\mu \rho }R^{\nu \sigma }=\frac{1}{12}%
I_{1}^{3}\ , \\
K_{2}& =A_{\mu \nu \rho \sigma }R^{\mu \rho }R^{\nu \sigma }=\frac{1}{36}%
I_{1}^{4}\ ,\quad K_{3}=C_{\mu \nu \rho \sigma }Q^{\mu \rho }Q^{\nu \sigma }=%
\frac{1}{48}I_{1}^{5}\ ,\quad K_{4}=A_{\mu \nu \rho \sigma }Q^{\mu \rho
}Q^{\nu \sigma }=\frac{1}{144}I_{1}^{6}\ , \\
K_{5}& =D_{\mu \nu \rho \sigma }Q^{\mu \rho }Q^{\nu \sigma }=0\ .
\end{align*}%
with 
\begin{gather*}
A_{\mu \nu \rho \sigma }=C_{\mu \nu \alpha \beta }C_{\gamma \delta \rho
\sigma }g^{\alpha \gamma }g^{\beta \delta }\ ,\quad \quad B_{\mu \nu \rho
\sigma }=C_{\mu \nu \alpha \beta }A_{\gamma \delta \rho \sigma }g^{\alpha
\gamma }g^{\beta \delta }\ , \\
D_{\mu \nu \rho \sigma }=B_{\mu \nu \rho \sigma }-\frac{J_{2}}{12}(g_{\mu
\rho }g_{\nu \sigma }-g_{\mu \sigma }g_{\nu \rho })-\frac{1}{4}J_{1}C_{\mu
\nu \rho \sigma }\ , \\
\tilde{D}_{\mu \nu \rho \sigma }=\frac{1}{\sqrt{J_{3}}}D_{\mu \nu \rho
\sigma }\ ,\quad \quad E_{\mu \nu \rho \sigma }=C_{\mu \nu \alpha \beta
}D_{\gamma \delta \rho \sigma }g^{\alpha \gamma }g^{\beta \delta }\ , \\
F_{\mu \nu \rho \sigma }=C_{\mu \nu \alpha \beta }E_{\gamma \delta \rho
\sigma }g^{\alpha \gamma }g^{\beta \delta }\ ,\quad \quad Q_{\mu }{}^{\nu
}=R_{\rho }{}^{\nu }R_{\mu }{}^{\rho }\ .
\end{gather*}%
The regularity condition at $X\rightarrow \pm \infty $ is the same as in Eq.
(\ref{cond2}). One may wonder if there exists a divergent higher
differential invariant of our space-time, such as $R_{;\mu \nu \cdots}
R^{;\mu \nu \cdots }$. A potential source of such divergences is the
presence of an anisotropic pressure that is not $C^{1}$ \cite{Musgrave:1995}%
. The energy-momentum tensor of the cosmic tube, however, is smooth
everywhere so that such a divergence cannot occur. Thus, \textit{all the
curvature invariants which can be built from the metric in Eqs. (\ref{WLP})
and (\ref{Rsol}) are regular everywhere} \textit{if the condition in Eq.} (%
\ref{cond2})\ \textit{on the integration constants holds}.

Interestingly enough, \textit{the function} $G(X,\theta )$ \textit{is not
relevant at all} as far as the regularity of the curvature invariants is
concerned. Thus, in particular, the curvature invariants do not depend on $%
\theta $. This implies that all the curvature invariants have a local
regular maximum at finite distance from the axis of the tube (as it will be
discussed in the next sections, in this radial coorinate $r$ the axis of the
tube is located at $X \rightarrow -\infty $ while the peaks of the curvature
invariants are at $X=0$ which is at finite distance from the axis). Since,
as it has been already emphasized, the curvature invariants do not depend on 
$\theta $, the regular peaks of the curvature invariants lie on a ring in
the $\left( X-\theta \right) $ plane. Such a ring (which is a tube from a
three-dimensional perspective) is also the support of the superconducting
currents associated with the present class of gravitating topologically
non-trivial solitons.

One can also verify that the space-time is of Petrov type \textbf{II}. One
more constraint on the integration constant $C_{2}$ will arise from the
analysis of the geodesic in the following sections.

\subsection{Periodicity of $U$ and the topological charge}

In this subsection we will discuss a simple but deep property of the $SU(2)$
valued scalar field $U$ which has a very important consequence. First of
all, one can notice that when 
\begin{equation}
q=\frac{1}{2}+n \ ,  \label{periodicity1}
\end{equation}%
with $n$ an integer, the topological charge is non-zero while, when $q$ is
an integer, the topological charge vanishes. This can be seen as follows.
The topological density corresponding to the NLSM configuration in Eq. (\ref%
{ansatz}) reads 
\begin{equation*}
\rho_{\text{B}}=\left( 12q\sin {q\theta }\sin ^{2}{\alpha }\right) \alpha
^{\prime }\ .
\end{equation*}%
Thus, as the range of $\theta $ is $\left[ 0,2\pi \right] $, the integral of
the above density%
\begin{equation*}
w_{\text{B}}=\int \rho_{\text{B}} dX \wedge d\theta \wedge dz \ ,
\end{equation*}
is non-zero if and only if Eq. (\ref{periodicity1}) holds. It is also worth
to note here that the $z $ coordinate goes along the axis of the tube and
along the topological density. Thus, in a sense, the quantity%
\begin{equation*}
\frac{w_{\text{B}}}{L_z}=\int \rho_{\text{B}} dX \wedge d\theta \ ,
\end{equation*}%
can be interpreted as the topological charge per unit of length of the tube,
with $L_z=\int dz$.

Here it is worth to emphasize that, when the condition in Eq. (\ref%
{periodicity1}) holds, $\theta $ is a proper angular coordinate with range $%
\left[ 0,2\pi \right] $. First of all, the metric itself is periodic with
period $2\pi $ as it depends on $\theta $ only through the function $%
G(X,\theta )$. The function $G(X,\theta )$\ depends\footnote{%
As it will be discussed in the next subsection, it is possible to choose the
integration constants of the solution in such a way to eliminate the deficit
angle close to the origin: see Eq. (\ref{Rat0}) and the discussion below.}
on $\theta $ only through the factor $\cos (2q\theta )$ in Eq. \eqref{EqG}.
Secondly, the energy-momentum tensor of the $SU(2)$-valued scalar field is
also periodic in $\theta $\ with the same period $2\pi $ when the condition
in Eq. (\ref{periodicity1}) holds.

However, the matrix $U$ itself is not periodic in $\theta $\ with the same
period $2\pi $ as it reads 
\begin{equation*}
U\ =\ 
\begin{pmatrix}
\cos {\alpha }+i\sin {\alpha }\cos (q\theta ) & ie^{-i(z+\omega t)}\sin {%
\alpha }\sin {q\theta } \\ 
ie^{i(z+\omega t)}\sin {\alpha }\sin (q\theta ) & \cos {\alpha }-i\sin {%
\alpha }\cos (q\theta )\ 
\end{pmatrix}%
\ .
\end{equation*}%
On the other hand, one can be sure that the \textquotedblleft lack of
periodicity\textquotedblright\ can be compensated by an internal Isospin
rotation since the energy-momentum tensor corresponding to the above NLSM
configuration is periodic with the correct $2\pi -$period: indeed, if such
\textquotedblleft lack of periodicity\textquotedblright\ could not be
compensated by an internal rotation, then the energy-momentum tensor would
not have the correct period. Consequently, $\theta $ is a proper angular
coordinate with range $\left[ 0,2\pi \right] $ when $q$ is half-integer.

Thus, as it happens with the spin-from-Isospin effect, the internal symmetry
group plays a fundamental role. In that case, the spin-from-Isospin effect
is generated by the possibility to require ``spherical symmetry up to an
internal rotation". In the present case, the condition that $U$ satisfies
periodic boundary conditions up to an Isospin rotation is enough to ensure
that $T_{\mu\nu}$ is periodic and $w_{\text{B}}\neq 0$.

\subsection{Coordinate transformation and the asymptotic behavior}

To analyze the nature of the metric in Eq. \eqref{WLP}, it is convenient to
make the following coordinate transformation: 
\begin{equation}
r(X)=\int_{-\infty }^{X}e^{-R(y)}dy=\int_{-\infty }^{X}\frac{%
e^{-C_{2}y-C_{3}}}{\big(\cosh (qy+C_{1})\big)^{K\kappa }}dy\ .
\label{coordtrans}
\end{equation}%
It is obvious that $r(X)$ increases monotonically as $X$ increases, since 
\begin{equation}
\frac{dr}{dX}=\frac{e^{-C_{2}X-C_{3}}}{\big(\cosh (qX+C_{1})\big)^{K\kappa }}%
>0\ .
\end{equation}%
Moreover, to make $r(X)$ well-defined, it is necessary to impose that 
\begin{equation}
C_{2} < K\kappa |q| \ .  \label{cond3}
\end{equation}
This assumption gives us the range of $r$ given by 
\begin{equation}
-\infty <X<\infty \qquad \Longrightarrow \qquad 0<r<\infty \ .
\end{equation}%
With this coordinate transformation, the metric becomes, 
\begin{equation*}
ds^{2}=-\frac{B_{0}^{2}\omega _{s}}{e^{f_{0}}}\Big(\frac{2e^{f_{0}}}{B_{0}}%
-\omega _{s}G\Big)dt^{2}-\frac{2B_{0}^{2}}{e^{f_{0}}}\Big(\frac{e^{f_{0}}}{%
B_{0}}-\omega _{s}G\Big)dtdz +\frac{B_{0}^{2}}{e^{f_{0}}}Gdz
^{2}+dr^{2}+e^{-2\tilde{R}(r)}d\theta ^{2}\ ,
\end{equation*}%
where $\tilde{R}(r)=(R\circ X)(r)$.

As we will show now, the function $e^{-2\tilde{R}(r)}$ of the new
cylindrical radial coordinate $r$ both for $r$ close to zero and for $%
r\rightarrow \infty $ is proportional to $r^{2}$:%
\begin{equation*}
e^{-2\tilde{R}(r)}\underset{r\rightarrow 0}{\rightarrow }\Delta _{0}r^{2}\ ,
\qquad e^{-2\tilde{R}(r)}\underset{r\rightarrow +\infty }{\rightarrow }%
\Delta _{\infty }r^{2}\ .\ 
\end{equation*}
This implies that $\theta $ is an angular coordinate. Consequently, it is
very important to determine the coefficients $\Delta _{0}$ and $\Delta
_{\infty }$ which determine the \textit{effective deficit angles} seen from
observers ``very close to" and ``very far from" the axis of the tube,
respectively.

Note that we have the following two limits of the function $R$; 
\begin{equation}
X \longrightarrow \pm \infty \qquad \Longrightarrow \qquad R \longrightarrow
(C_{2} \pm K \kappa |q|) X \ ,
\end{equation}
so that we also have 
\begin{equation}
X \longrightarrow \pm \infty \qquad \Longrightarrow \qquad e^{-R}
\longrightarrow 2^{K\kappa} e^{\mp K\kappa C_{1} - C_{3}} e^{-(C_{2} \pm
K\kappa |q|) X} \ .
\end{equation}
In the limit of $X \gg X_{+}$ for a sufficiently large $X_{+}\gg 0$, we find
that 
\begin{eqnarray}
&& r(X) = r_{1} + \frac{ 2^{K\kappa} \times e^{-(K\kappa C_{1}+C_{3})}} {%
-(C_{2}+K\kappa |q|)} \Big( e^{-(C_{2}+K\kappa |q|) X } - e^{-(C_{2}+K\kappa
|q|) X_{+} }\Big)  \notag \\
&& \hspace{.35in} \approx \frac{ 2^{K\kappa} \times e^{-(K\kappa
C_{1}+C_{3})}} {-(C_{2}+K\kappa |q|)} \times e^{-(C_{2}+K\kappa |q|) X} \ .
\end{eqnarray}
Here, we have defined a finite constant 
\begin{equation}
r_{1} \equiv \int_{-\infty}^{X_{+}} \frac{e^{-C_{2} y -C_{3}}} {\big( %
\cosh(|q|y+C_{1})\big)^{K\kappa}} dy \ ,
\end{equation}
and in the last line, we used the assumption that $C_{2} + K\kappa |q| < 0$.
Thus, we obtain 
\begin{equation}
e^{-2R} \approx \big( K\kappa |q| + C_{2} \big)^2 r^{2}\ , \qquad \qquad 
\text{at}\quad X \gg 1 \ .
\end{equation}
In a similar way, the limit of $X \ll X_{-}$ for a sufficiently small $X_{-}
\ll 0$ is found to be 
\begin{eqnarray}
&& r(X) = r_{2} - \frac{ 2^{K\kappa} \times e^{K\kappa C_{1}-C_{3}}} {%
-(C_{2}-K\kappa |q|)} \Big( e^{-(C_{2}-K\kappa |q|) X_{-} } -
e^{-(C_{2}-K\kappa |q|) X }\Big)  \notag \\
&& \hspace{.35in} \approx \frac{ 2^{K\kappa} \times e^{K\kappa C_{1}-C_{3}}%
} {-(C_{2}-K\kappa |q|)} \times e^{-(C_{2}-K\kappa |q|) X } \ .
\end{eqnarray}
Here, we also have defined a finite constant 
\begin{equation}
r_{2} \equiv \int_{-\infty}^{X_{-}} \frac{e^{-C_{2} y -C_{3}}} {\big( %
\cosh(|q|y+C_{1})\big)^{K\kappa}} dy \ ,
\end{equation}
and in the last line we have used the fact that $X \ll X_{-} \ll 0$.
Therefore, we get 
\begin{equation}  \label{Rat0}
e^{-2R} \approx \big( K\kappa |q| - C_{2} \big)^2 r^{2} e^{-2 C_3}\ , \qquad
\qquad \text{at}\quad X \ll 0 \ .
\end{equation}
Now, we see that we can choose $C_3$ such that the angular deficit is $1$
near the axis defined by $r(X) =0$ so that $\theta$ becomes a proper angular
coordinate. The ratio of the values of $g_{\theta \theta}$ at two infinities
becomes, 
\begin{equation}
\frac{g_{\theta \theta}(X=\infty)}{g_{\theta \theta}(X=-\infty)} = \frac{%
g_{\theta \theta}(r=\infty)}{g_{\theta \theta}(r=0)} = \Big( \frac{K\kappa
|q| + C_{2}}{K\kappa |q| - C_{2}} \Big)^{2} < 1 \ .
\end{equation}
When we assume that $G$ dies out at spatial infinity, as supported by the
plot given in Fig. \ref{fig:PlotG}, the asymptotic form of the metric is
found to be 
\begin{align}
ds^{2} = -\frac{B_{0}}{2 \omega_{s}} \big( 2 \omega_{s} dt + dz \big)^{2} + 
\frac{B_{0}}{2 \omega_{s}} dz^{2} + dr^{2} + \Big( \frac{K\kappa |q| + C_{2}%
}{K\kappa |q| - C_{2}} \Big)^{2} r^{2} d\theta^{2} \ ,
\end{align}
which describes a boosted cosmic string at spatial infinity.

\subsection{Geodesics}

Let $\lambda $ be an affine parameter of geodesics in our space-time. The
geodesic equations are found to be 
\begin{eqnarray}
&&\hspace{-0.2in}\ddot{t}-B_{0}e^{-f_{0}}\big(\omega _{s}\dot{t}+\dot{z}\big)%
\frac{dG}{d\lambda }=0\ ,  \label{geodesict} \\
&&\hspace{-0.2in}\ddot{z}+\omega _{s}B_{0}e^{-f_{0}}\big(\omega _{s}\dot{t}+%
\dot{z}\big)\frac{dG}{d\lambda }=0\ ,  \label{geodesicphi} \\
&&\hspace{-0.2in}\ddot{X}-\big(\dot{X}^{2}-\dot{\theta}^{2}\big)\big(%
C_{2}+K\kappa q\tanh (qX+C_{1})\big)  \notag \\
&&-\frac{1}{2}B_{0}^{2}\big(\omega _{s}\dot{t}+\dot{z}\big)%
^{2}e^{2C_{2}X+2C_{3}-f_{0}}\cosh ^{2K\kappa }(qX+C_{1})\partial _{X}G=0\ ,
\label{geodesicr} \\
&&\hspace{-0.2in}\ddot{\theta}-2\dot{X}\dot{\theta}\big(C_{2}+K\kappa q\tanh
(qX+C_{1})\big)  \notag \\
&&-\frac{1}{2}B_{0}^{2}\big(\omega _{s}\dot{t}+\dot{z}\big)%
^{2}e^{2C_{2}X+2C_{3}-f_{0}}\cosh ^{2K\kappa }(qX+C_{1})\partial _{\theta
}G=0\ ,  \label{geodesictheta}
\end{eqnarray}%
where 
\begin{equation}
\frac{dG}{d\lambda }=\dot{X}\partial _{X}G+\dot{\theta}\partial _{\theta }G\
,
\end{equation}%
and the dot denotes the derivative with respect to $\lambda $.

From the first two equations, we obtain 
\begin{gather}
\omega_{s} \ddot{t} + \ddot{z} = 0 \qquad \Longrightarrow \qquad \omega_{s}
t + z = a \lambda + b \ , \\
\ddot{z}-\omega_s \ddot{t}+2B_0 e^{-f_0} \omega_s (\omega_s \dot{t}+\dot{z})%
\frac{dG}{d\lambda} = 0 \ ,
\end{gather}
for some constants $a$ and $b$.

Then, Eqs. (\ref{geodesicphi}), (\ref{geodesicr}), and (\ref{geodesictheta})
become 
\begin{eqnarray}
&& \hspace{-.2in} \ddot{z}+aB_0e^{-f_0}\omega \frac{dG}{d\lambda} =0 \ ,
\label{geodesicphi1} \\
&& \hspace{-.2in} \ddot{X} - \big( \dot{X}^{2} - \dot{\theta}^{2} \big) %
\big( C_{2} + K\kappa q \tanh(qX+C_{1}) \big)  \notag \\
&&- \frac{1}{2} B_{0}^{2} a^2 e^{2C_{2}X+2C_{3}- f_{0}} \cosh^{2 K\kappa}
(qX+C_{1}) \partial_{X}G= 0 \ ,  \label{geodesicr1} \\
&&\hspace{-.2in} \ddot{\theta} - 2 \dot{X} \dot{\theta} \big( C_{2} +
K\kappa q \tanh(qX+C_{1}) \big)  \notag \\
&& - \frac{1}{2} B_{0}^{2} a^{2} e^{2C_{2}r+2C_{3}- f_{0}} \cosh^{2 K\kappa}
(qX+C_{1}) \partial_{\theta}G= 0 \ .  \label{geodesictheta1}
\end{eqnarray}

On the hypersurface of constant $z$, the Eq. (\ref{geodesicphi}) becomes 
\begin{equation}
\omega_{s} B_{0} e^{-f_{0}} \omega_{s} \dot{t} \ \frac{d G}{d\lambda} = 0 \ ,
\end{equation}
from which we obtain 
\begin{equation}
\frac{d G}{d\lambda} = \dot{X} \partial_{X}G + \dot{\theta}
\partial_{\theta}G = 0 \ .  \label{zeroG}
\end{equation}
Then, Eq. (\ref{geodesict}) becomes 
\begin{equation}
\ddot{t} = 0 \qquad \Longrightarrow \qquad t = \frac{a}{\omega_{s}} \lambda
+ b \ .
\end{equation}
For convenience, let's put 
\begin{equation}
a = \omega_{s} \ , \qquad b = 0\ , \qquad \text{so that} \qquad t = \lambda
\ .
\end{equation}
Then, the remaining geodesic equations become 
\begin{eqnarray}
&& \hspace{-.2in} \ddot{X} - \big( \dot{X}^{2} - \dot{\theta}^{2} \big) %
\big( C_{2} + K\kappa q \tanh(qX+C_{1}) \big)  \notag \\
&&- \frac{1}{2} B_{0}^{2} \ \omega_{s}^2 e^{2C_{2}X+2C_{3}- f_{0}}
\cosh^{2K\kappa} (qX+C_{1}) \partial_{X}G= 0 \ ,  \label{geodesicr2} \\
&& \hspace{-.2in} \ddot{\theta} - 2 \dot{X}\dot{\theta} \big( C_{2} +
K\kappa q \tanh(qX+C_{1}) \big)  \notag \\
&&- \frac{1}{2} B_{0}^{2} \omega_{s}^{2} e^{2C_{2}X+2C_{3}- f_{0}} \cosh^{2
K\kappa} (qX+C_{1}) \partial_{\theta}G= 0 \ .  \label{geodesictheta2}
\end{eqnarray}
\noindent Using Eq. (\ref{zeroG}) in a linear combination $\Big(\dot{X}%
\times (\ref{geodesicr1})+\dot{\theta}\times (\ref{geodesictheta1})\Big)$
yields 
\begin{equation*}
\frac{1}{2}\frac{d}{d\lambda }\big(\dot{X}^{2}+\dot{\theta}^{2}\big)-\dot{X}%
\big(\dot{X}^{2}+\dot{\theta}^{2}\big)\big(C_{2}+K\kappa q\tanh (qX+C_{1})%
\big)=0\ ,
\end{equation*}%
or equivalently, 
\begin{equation}
\frac{d}{d\lambda }\Big[\frac{1}{2}\ln \big(\dot{X}^{2}+\dot{\theta}^{2}\big)%
-C_{2}X-K\kappa \ln \big\{\cosh (qX+C_{1})\big\}\Big]=0\ .
\end{equation}%
Thus, for some constant $C_{4}$, we have 
\begin{equation}
\dot{X}^{2}+\dot{\theta}^{2}=e^{2(C_{2}X+C_{4})}\cosh ^{2K\kappa
}(qX+C_{1})\ .  \label{intg}
\end{equation}%
We should assume that 
\begin{equation}
C_{2}<-K\kappa \ |q|\ ,  \label{cond4}
\end{equation}%
to avoid infinity velocities at large $X$. \noindent The line element for
the geodesic with constant $z $ is 
\begin{equation*}
ds^{2}=-\frac{B_{0}^{2}\omega _{s}}{e^{f_{0}}}\Big(\frac{2e^{f_{0}}}{B_{0}}%
-\omega _{s}G\Big)dt^{2}+e^{-2R}\big(dX^{2}+d\theta ^{2}\big)\ .
\end{equation*}%
Divided by the affine parameter $t$, it can be written as 
\begin{equation*}
B_{0}\omega _{s}\Big(B_{0}\omega _{s}e^{-f_{0}}G-2\Big)+e^{-2R}\big(\dot{X}%
^{2}+\dot{\theta}^{2}\big)=\frac{ds^{2}}{dt^{2}}\equiv \epsilon \ .
\end{equation*}%
By an appropriate rescaling $\epsilon $ becomes $-1$ or $0$ for time-like or
null geodesic, respectively. Using Eq. (\ref{intg}), we have 
\begin{equation*}
e^{-2R}\big(\dot{X}^{2}+\dot{\theta}^{2}\big)=e^{-2C_{2}X-2C_{3}}\sech%
^{2K\kappa }\big(qX+C_{1}\big)\times e^{2(C_{2}X+C_{4})}\cosh ^{2K\kappa
}(qX+C_{1})=e^{-2C_{3}+2C_{4}}\ .
\end{equation*}%
Thus, the line element of the geodesic with constant $z $ gives a relation
between constants as follows 
\begin{equation*}
B_{0}\omega _{s}\Big(B_{0}\omega _{s}e^{-f_{0}}G-2\Big)+e^{-2C_{3}+2C_{4}}-%
\epsilon =0\ .
\end{equation*}%
Using Eq. (\ref{intg}) in the geodesic equation of $r$, from Eq. (\ref%
{geodesicr2}) one finds that 
\begin{align}
\ddot{X} & -2\dot{X}^{2}\big(C_{2}+K\kappa q\tanh (qX+C_{1})\big)%
-e^{2C_{2}X} \cosh ^{2K\kappa }(qX+C_{1})  \notag \\
& \times \Big\{e^{2C_{4}}\big(C_{2}+K\kappa q\tanh (qX+C_{1})\big)+\frac{1}{2%
}B_{0}^{2}\omega _{s}^{2}e^{2C_{3}-f_{0}}\partial _{X}G\Big\}=0\ .
\end{align}%
After solving this equation, one should plug the solution to the equation of
motion for $\theta $ in Eq. (\ref{geodesictheta2}) to find $\theta (t)$.
This completes the solving process of the geodesic equations with constant $%
z $. In principle, this problem can be reduced to an effective
one-dimensional Newtonian problem observing that, along the geodesics with
constant $z $, one has%
\begin{equation}  \label{aux2}
\dot{X}\partial _{X}G+\dot{\theta}\partial _{\theta }G=0\ \ \Rightarrow \ \ 
\frac{d\theta }{dr}=-\frac{\partial _{r}G}{\partial _{\theta }G}\ ,
\end{equation}%
where $G(X,\theta )$ is defined in Eqs. (\ref{splitG}), (\ref{finalG0}) and (%
\ref{finalG}). The reason is that from Eq. (\ref{aux2}) one can determine $%
\theta =\theta \left( X \right) $ along the geodesics with constant $z $.
Once $\theta =\theta \left( X \right) $ has been determined, one can insert
it into Eq. (\ref{intg}) obtaining a first order Newtonian-like equation of
the form%
\begin{eqnarray*}
\dot{X}^{2} &=&V_{\text{eff}}\left( X \right) \ , \\
V_{\text{eff}}\left( X \right) &=&\frac{e^{2(C_{2}X+C_{4})}\cosh ^{2K\kappa
}(qX+C_{1})}{1+\left( \frac{d\theta }{dX}\right) ^{2}}\ .
\end{eqnarray*}%
However, Eq. (\ref{aux2}) is a quite complicated first order non-autonomous
differential equation for $\theta \left( X \right) $ due to the explicit
form of $G(X,\theta )$ in Eqs. (\ref{splitG}), (\ref{finalG0}) and (\ref%
{finalG}). The relevant issue of the behavior of geodesics in this family of
gravitating solitons deserves a more detailed analysis on which we hope to
come back in a future publication.

In the usual case of cosmic strings, the angle defect of gravitational
lensing is independent of the initial distance of the particle from the
source. But in our case, the distribution of source is smoothly spread out
to $X=\infty $, so that the angle defect depends on $X_{0}$, the initial
location of geodesic motion. This would be one of the most distinguished
properties of our solution.

\subsection{Constraint on integration constant}

Combining the regularity conditions on the integration constants in Eqs. %
\eqref{cond1}, \eqref{cond2}, \eqref{cond3} and \eqref{cond4} we obtain the
following single inequality 
\begin{equation}  \label{condfinal}
-(1-K\kappa )\ |q|<C_{2}<-K\kappa \ |q|\ .
\end{equation}
Thus, when $C_2$ satisfies the above condition and $C_3$ is chosen as in Eq. %
\eqref{Rat0} all the curvature invariants of the metric are regular and the
geodesics behave in a reasonable way. Note that the experimental value of
the Pions coupling constant is such that $0<K\kappa\ll 1$. On the other
hand, the integration constant $C_{1}$ is a free parameter which fixes the
location of the point about which the profile function $\alpha(X)$ is
symmetric.

\section{Gauged gravitating tubes}

In this Section we promote our configurations to be solutions of the
Einstein-Maxwell-NLSM system, and also show how these are superconducting
gravitating tubes.

\subsection{The Einstein-Maxwell-NLSM theory}

The action of the $U(1)$ gauged Einsten NLSM theory is 
\begin{gather}
I[g,U,A]=\int d^{4}x\sqrt{-g}\left[ \frac{\mathcal{R}}{2\kappa }+\frac{K}{4}%
\mathrm{Tr}\left( L^{\mu }L_{\mu }\right) -\frac{1}{4}F_{\mu \nu }F^{\mu \nu
}\right] \ ,  \label{sky1} \\
L_{\mu }=U^{-1}D_{\mu }U\ ,\ \ \ D_{\mu }=\nabla _{\mu }+A_{\mu }\left[
t_{3},\ .\ \right] \ , \ F_{\mu \nu }=\partial _{\mu }A_{\nu }-\partial
_{\nu }A_{\mu }\ .  \label{sky2.5}
\end{gather}%
The respective field equations are 
\begin{equation}
D_{\mu }L^{\mu }=0\ ,\qquad G_{\mu \nu }=\kappa (T_{\mu \nu }+\bar{T}_{\mu
\nu })\ ,
\end{equation}%
\begin{equation}
\nabla _{\mu }F^{\mu \nu }=J^{\nu }\ ,  \label{maxwellNLSM}
\end{equation}%
where the current $J^{\mu }$ is given by 
\begin{equation}
J^{\mu }=\frac{K}{2}\text{Tr}\left[ \widehat{O}L^{\mu }\right] \ ,\qquad 
\widehat{O}=U^{-1}t_{3}U-t_{3}\ ,  \label{current}
\end{equation}%
and $\bar{T}_{\mu \nu }$ is the energy-momentum tensor of the Maxwell theory 
\begin{equation}
\bar{T}_{\mu \nu }=F_{\mu \alpha }F_{\nu }^{\;\alpha }-\frac{1}{4}F_{\alpha
\beta }F^{\alpha \beta }g_{\mu \nu } \ .
\end{equation}%
Note that in the action in Eq. \eqref{sky1} there is a quadratic term in $%
A_\mu$.

The topological charge in this case \cite{Callan} is given by 
\begin{equation}
w_{\text{B}}=\frac{1}{24\pi^{2}}\int_{\Sigma}\rho_{\text{B}}\ ,
\end{equation}
where 
\begin{equation}  \label{rhoBCW}
\rho_{\text{B}}=\epsilon^{ijk}\text{Tr}\biggl[\left(
U^{-1}\partial_{i}U\right) \left( U^{-1}\partial_{j}U\right) \left(
U^{-1}\partial_{k}U\right) -\partial_{i}\left[ 3A_{j}t_{3}\left(
U^{-1}\partial_{k}U+\left( \partial_{k}U\right) U^{-1}\right) \right] \biggl]%
\ .
\end{equation}
Note that the second term in Eq. \eqref{rhoBCW}, the Callan-Witten term,
guarantees both the conservation and the gauge invariance of the topological
charge.

\subsection{Gravitating tubes coupled to the electromagnetic field}

Let's consider the following Maxwell potential 
\begin{equation}
A_{\mu }=\left( u,0,0,\frac{1}{\omega _{s}}u\right) \ ,\qquad u=u(X,\theta
)\ ,  \label{Amu}
\end{equation}%
together with the metric defined in Eq. \eqref{WLP} and the matter field in
Eq. \eqref{ansatz}. This is a very convenient ansatz \cite{crystal1}, \cite%
{crystal2}, because a direct computation reveals that the three field
equations for the gauged NLSM reduce (once again) to 
\begin{equation*}
\alpha ^{\prime \prime }-\frac{q^{2}}{2}\sin {2\alpha }=0\ ,
\end{equation*}%
so that 
\begin{equation}
\alpha (X)=2\arctan (e^{qX+C_{1}})\ .  \label{KK0}
\end{equation}%
On the other hand, the four Maxwell equations reduce to just one linear
equation: 
\begin{equation}
\Delta u-2Ke^{-2R}(\omega _{s}-2u)\sin ^{2}{\alpha }\sin ^{2}{q\theta }=0\ .
\label{maxremaining}
\end{equation}%
Note that this linear equation for $\Psi =\omega _{s}-2u$ can be easily
solved, at least numerically, since $\alpha (X)$ is explicitly known and $%
R(X)$ can be also determined explicitly. In fact, there are two non-trivial
Einstein equations that can be combined to obtain an uncoupled equation for $%
R$, namely 
\begin{equation*}
R^{\prime \prime }-\frac{1}{2}K\kappa (\alpha ^{\prime 2}+q^{2}\sin ^{2}{%
\alpha })=0\ ,
\end{equation*}%
so that, as in the case without Maxwell, 
\begin{equation}
R(X)=K\kappa \log (\cosh (qX+C_{1}))+C_{2}X+C_{3}\ ,  \label{R}
\end{equation}%
and $G$ satisfies the following equation 
\begin{equation}
\Delta G+\frac{2\kappa }{B_{0}^{2}\omega _{s}^{2}}e^{f_{0}-2R}\biggl(K\sin
^{2}{\alpha }\sin ^{2}{q\theta }(\omega _{s}-2u)^{2}+e^{2R}(\nabla u)^{2}%
\biggl)=0\ .  \label{KK}
\end{equation}%
Hence, once again, the function $G$ satisfy a flat two-dimensional Poisson
equation in which the source is explicitly known. Therefore, once the
Maxwell equation in Eq. (\ref{maxremaining}) has been solved, the function $%
G $ can be determined explicitly using several methods (such as the Green
functions).

Resuming, using the ansatz for $U$ and $A_{\mu }$ in Eqs. \eqref{ansatz}, %
\eqref{Amu} and for the metric in Eq. \eqref{WLP}, the Einstein-Maxwell-NLSM
field equations reduce simply to Eqs. \eqref{maxremaining} and \eqref{KK},
where $\alpha (X)$ and $R(X)$ are in Eqs. \eqref{KK0} and \eqref{R}. It is
worth to emphasize that this is a quite remarkable reduction of such a
coupled system of non-linear PDEs in a sector with non-trivial topological
charge as the NLSM (which is already a very complicated theory in itself) is
coupled directly both to GR and to the $U(1)$ gauge field $A_{\mu }$.

The energy density measured in an orthonormal frame is found to be 
\begin{align}
T_{\hat{0}\hat{0}} =& \frac{K e^{f_{0}}}{2 \omega_{s} B_{0} (2 e^{f_{0}} -
B_{0}\omega_{s} G )} \Biggl[ 2\sin^{2}(q\theta) \sech^{2}(qX+C_{1}) (
2u-\omega_{s} )^{2}  \notag \\
& + e^{2C_{2}X + C_{3}} \cosh^{2K\kappa} (qX+ C_{1}) \Big\{ %
(\partial_{X}u)^{2} + (\partial_{\theta}u)^{2} \Big\} \Biggl]  \notag \\
&+ K q^{2} e^{2C_{2}X + C_{3}} \sech^{2(1-K\kappa)}(qX+C_{1}) \ ,
\end{align}
where we used the tetrad given by 
\begin{align}
\hat{\mathbf{e}}_{0} = \frac{1}{\sqrt{-g_{00}}} \partial_{t} \ .  \notag
\end{align}
The topological charge density, including the Callan-Witten term, is 
\begin{equation*}
\rho _{\text{B}}=\partial _{X}\Big(6q\big(\alpha -\sin \alpha \cos \alpha %
\big)\sin (q\theta )+\frac{12q}{\omega _{s}}\sin \alpha \cos \alpha \sin
(q\theta )\cdot u\Big)+\partial _{\theta }\Big(\frac{12}{\omega _{s}}\cos
(q\theta )\cdot u\partial _{X}\alpha \Big)\ .
\end{equation*}%
The current is given by 
\begin{equation}
J_{\mu}=2K\sin ^{2}{\alpha }\sin ^{2}{q\theta }(\partial_\mu \Phi-2 A_\mu)\ ,
\label{currgaugedtubes}
\end{equation}%
and the components of the electric and magnetic fields read 
\begin{equation*}
E_{X}=-\partial _{X}u\ ,\qquad E_{\theta }=-\partial _{\theta }u\ ,\qquad
B_{X}=-\frac{1}{B_{0}}e^{2R}\partial _{\theta }u\ ,\qquad B_{\theta }=\frac{1%
}{B_{0}}e^{2R}\partial _{X}u\ .
\end{equation*}
The plots here below as well as the review of the Witten construction
clearly show why these solutions of the Einsten-Maxwell-NLSM represent
regular topologically non-trivial gravitating superconducting tubes.

In order to plot the relevant physical functions we have set our parameters
as 
\begin{gather}
C_1=0\ , \quad C_2=-\frac{1}{50}\ ,\quad C_3=0\ , \quad K=\frac{1}{10}\
,\quad \kappa=\frac{1}{4}\ ,  \notag \\
q=\frac{1}{2}\ ,\quad \omega_s=1\ , \quad f_0=0\ ,\quad B_0=\frac{1}{2} \ .
\end{gather}

\begin{figure}[h!]
\centering
\includegraphics[scale=.45]{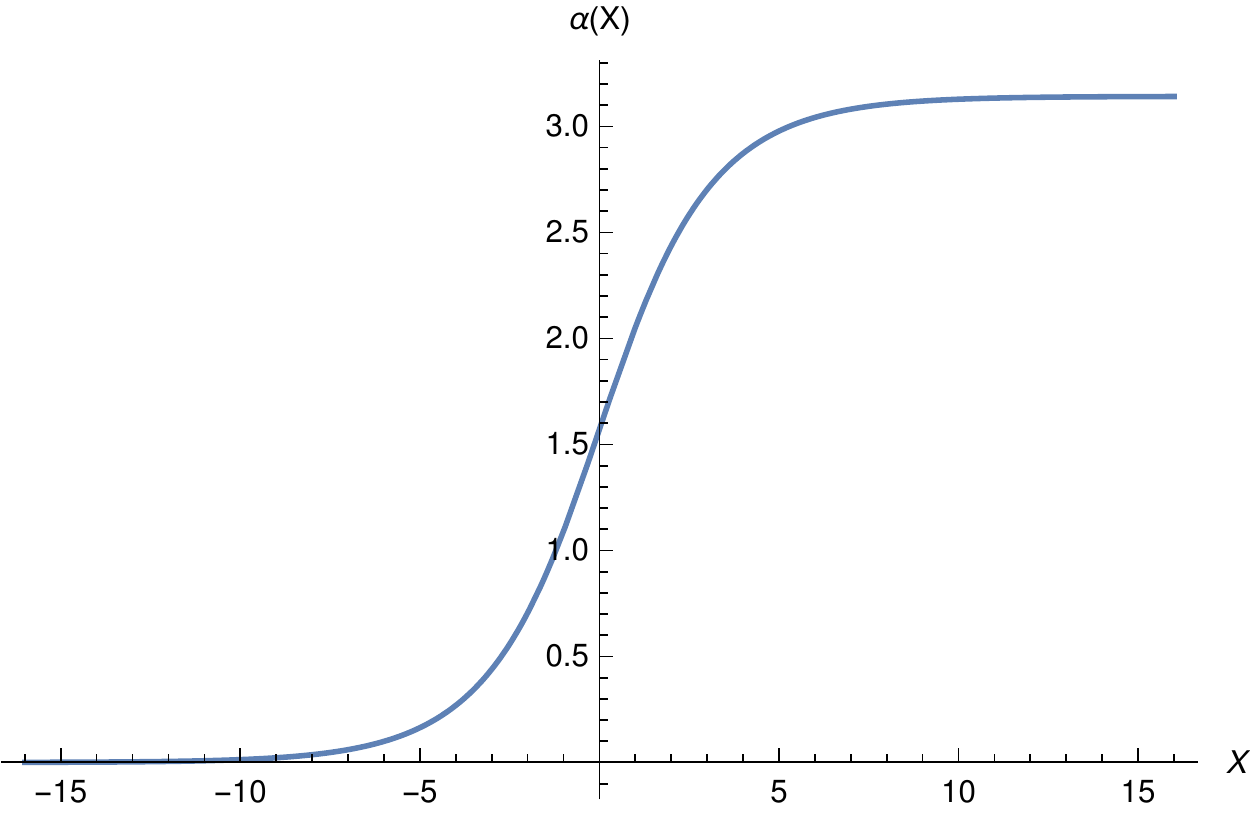}\ \ %
\includegraphics[scale=.45]{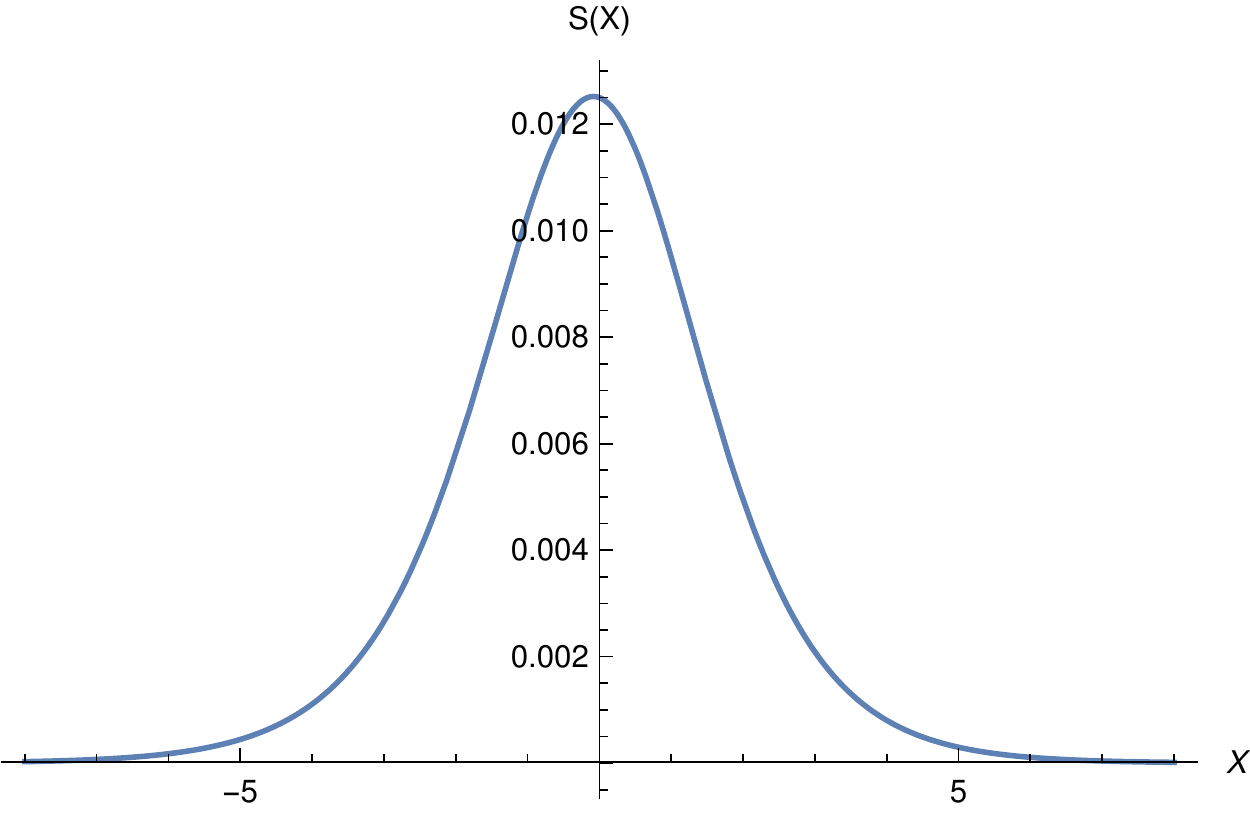}\ \ %
\includegraphics[scale=.45]{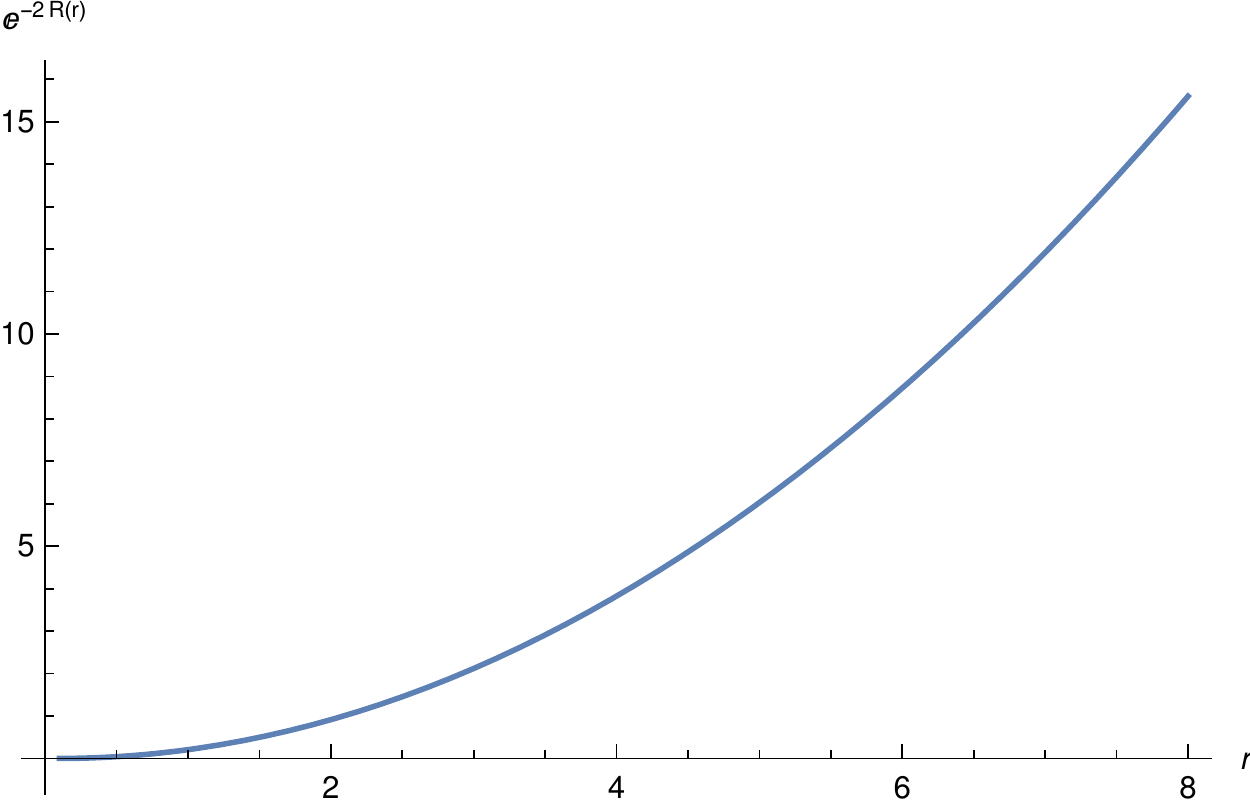}
\caption{From left to right, the $\protect\alpha$ profile and the Ricci
scalar $S$ as functions of the $X$ coordinate and metric function $%
e^{-2R(r)} $ as a function of the $r$ coordinate.}
\label{fig:scalars}
\end{figure}

\begin{figure}[h!]
\centering
\includegraphics[scale=.6]{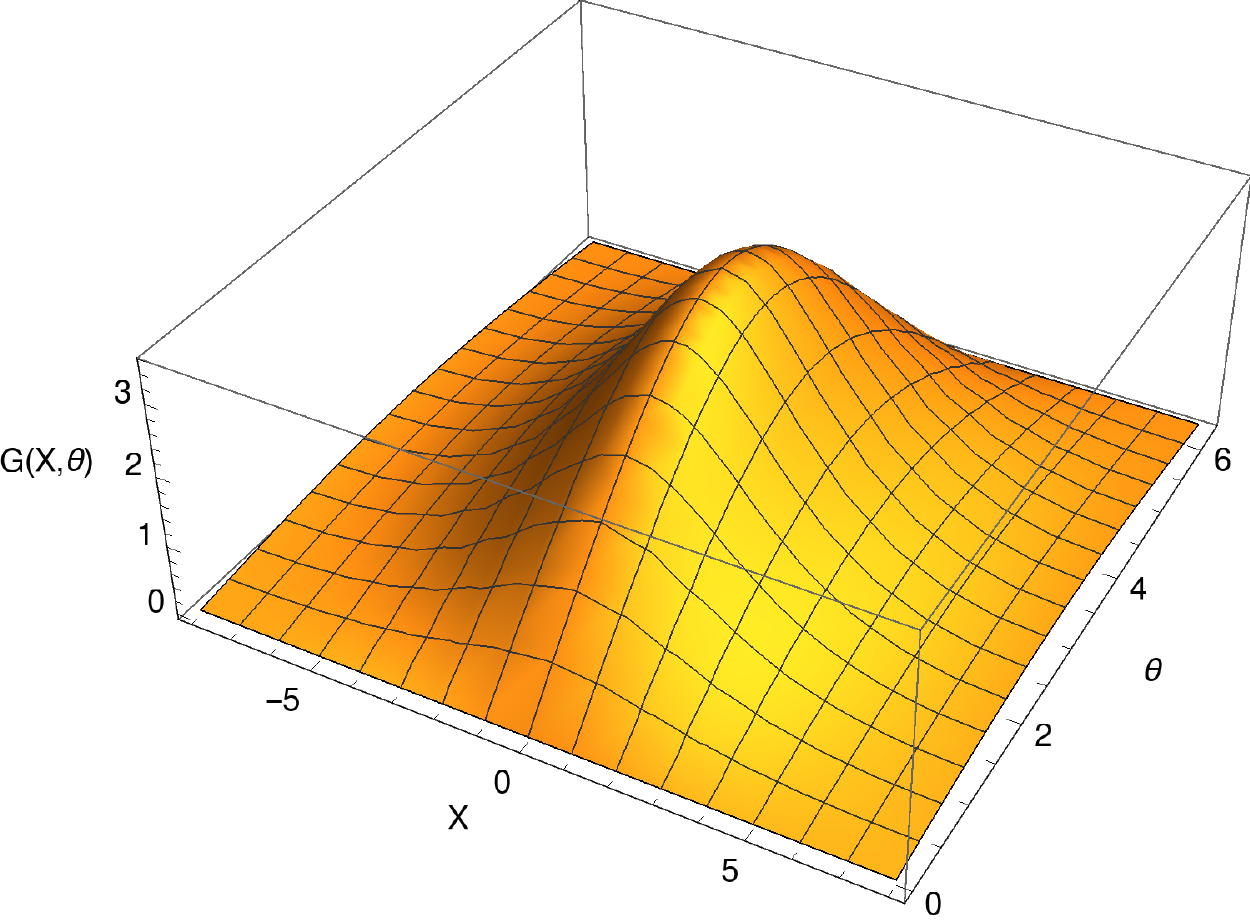}
\caption{The metric function $G$ as a function of $X$ and $\protect\theta$.}
\label{fig:PlotG}
\end{figure}

From Fig. \ref{fig:scalars} and Fig. \ref{fig:PlotG} we can see that even if
the metric is not axi-symmetric due to the explicit $\theta$-dependence of $%
G(X,\theta)$ in Eq. \eqref{KK}, all the curvature invariants are
axi-symmetric as they only depend on $X$. Thus, the plots of all the
curvature invariants are very similar and they all show a smooth peak at
finite distance from the origin (remember that, in the coordinate $X$, the
origin is at $X\rightarrow -\infty$). As we expected, $e^{-2R(r)}$ goes as $%
r^2$ when $r\rightarrow 0$ as well as when $r\rightarrow\infty$.

\begin{figure}[h!]
\centering
\includegraphics[scale=.55]{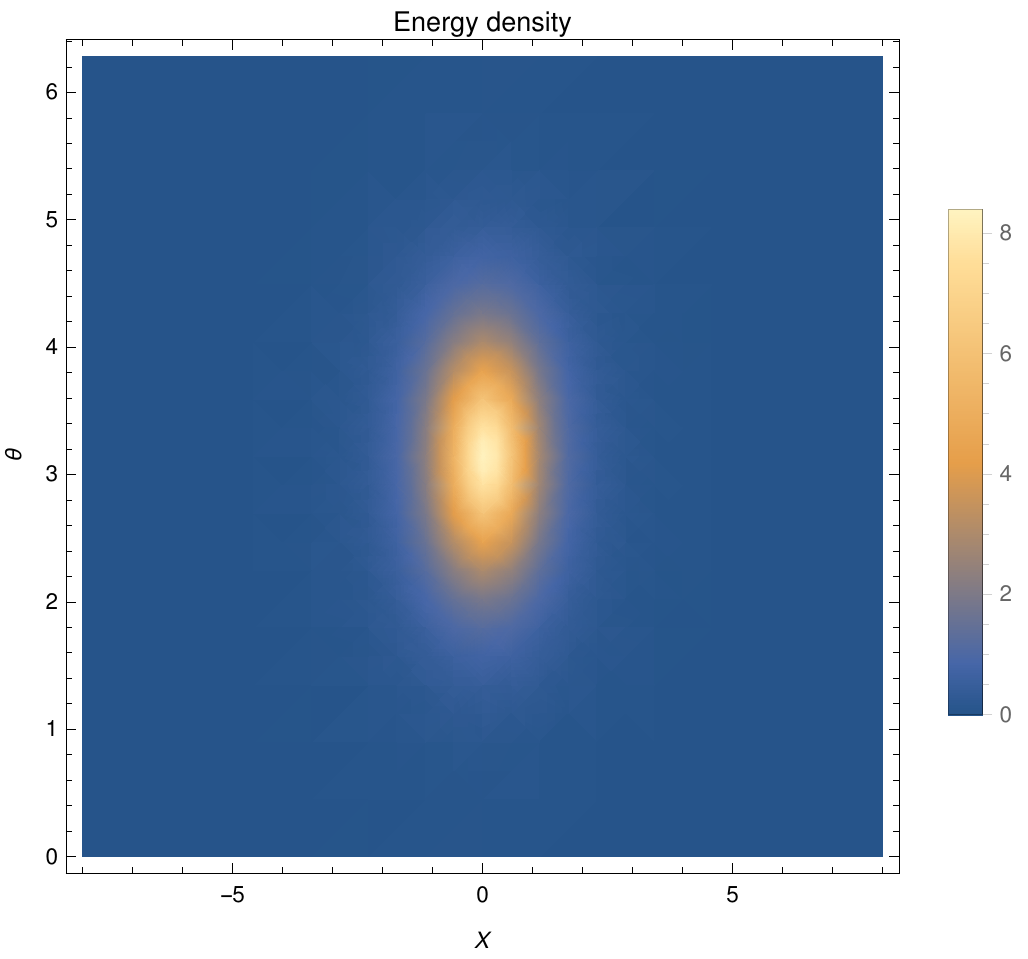} \ \ %
\includegraphics[scale=.55]{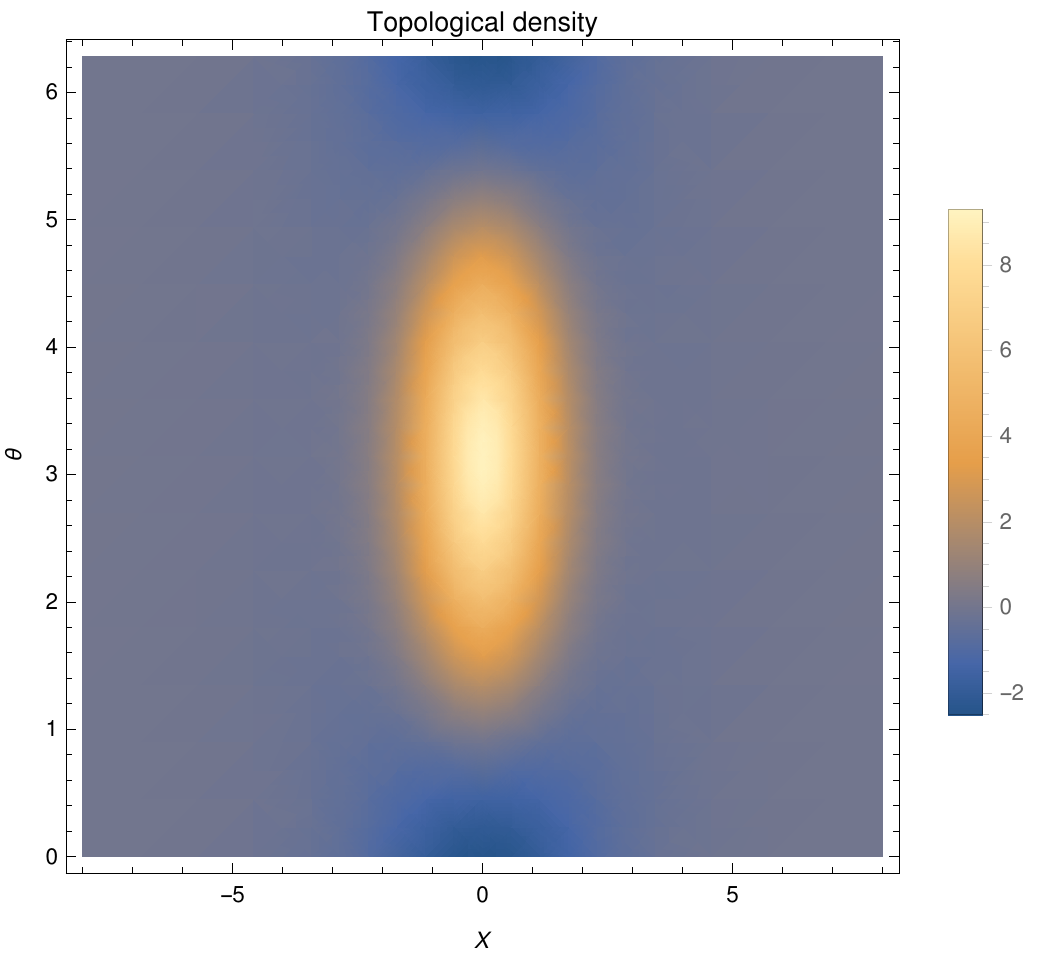}
\caption{The Energy density $T_{\hat{0}\hat{0}}$ and topological density $%
\protect\rho_{\text{B}}$ as functions of $X$ and $\protect\theta$.}
\label{fig:densities}
\end{figure}

In Fig. \ref{fig:densities}, the energy density associated to a comoving
observer has two parts. The first one only depends on ``$X$" and, as the
curvature invariants, has its smooth maximum at the same finite dinstance
from the axis. The second part depends both on ``$X$" and on ``$\theta$" and
the corresponding peak is at the same distance from the axis as the peak of
the first term and, in $\theta$ is localized around $\theta\sim \pi$. The
peaks of the energy density and the peaks of the topological density
coincide, as expected.

\begin{figure}[h!]
\centering
\includegraphics[scale=.4]{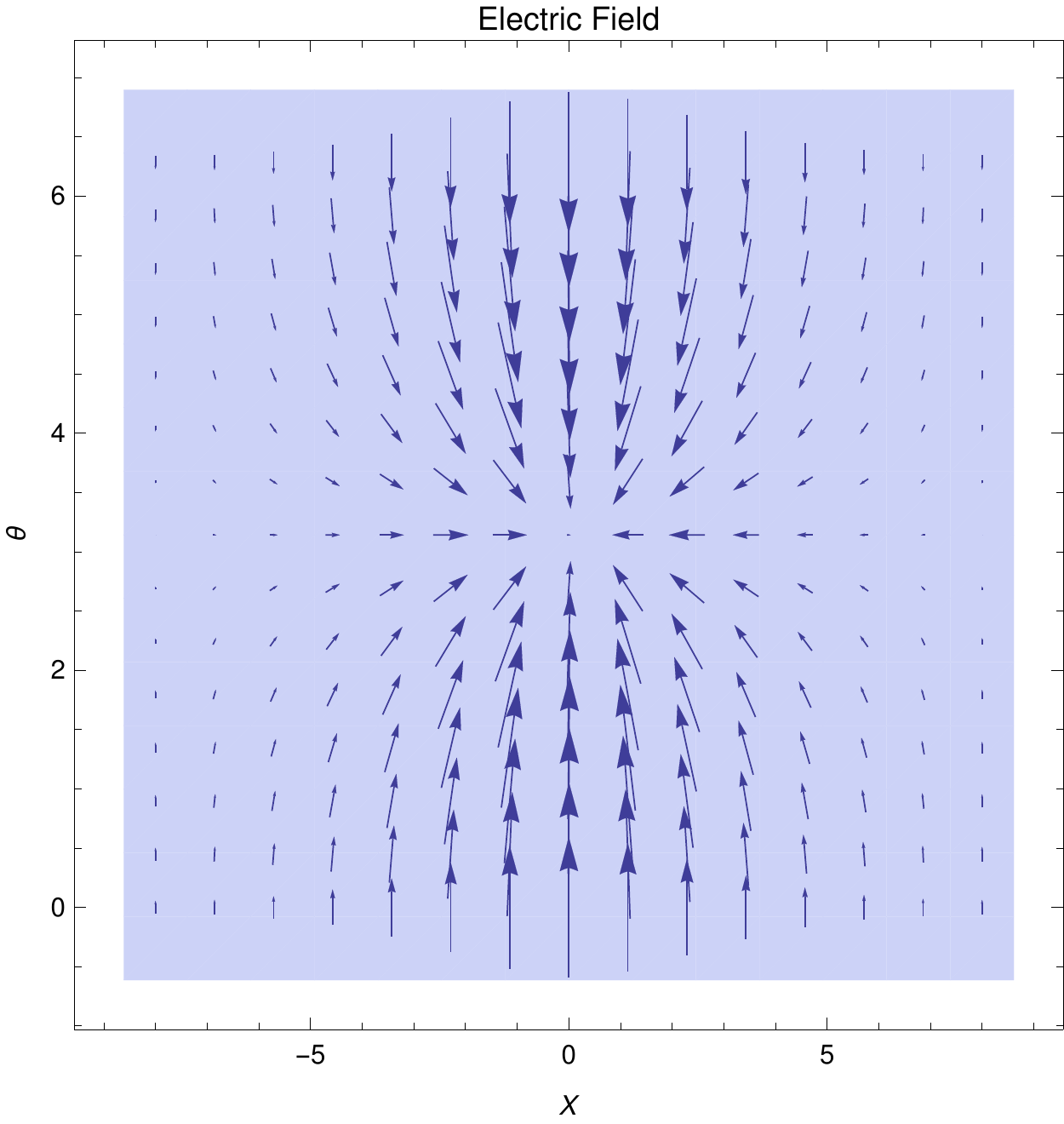}\ \ %
\includegraphics[scale=.4]{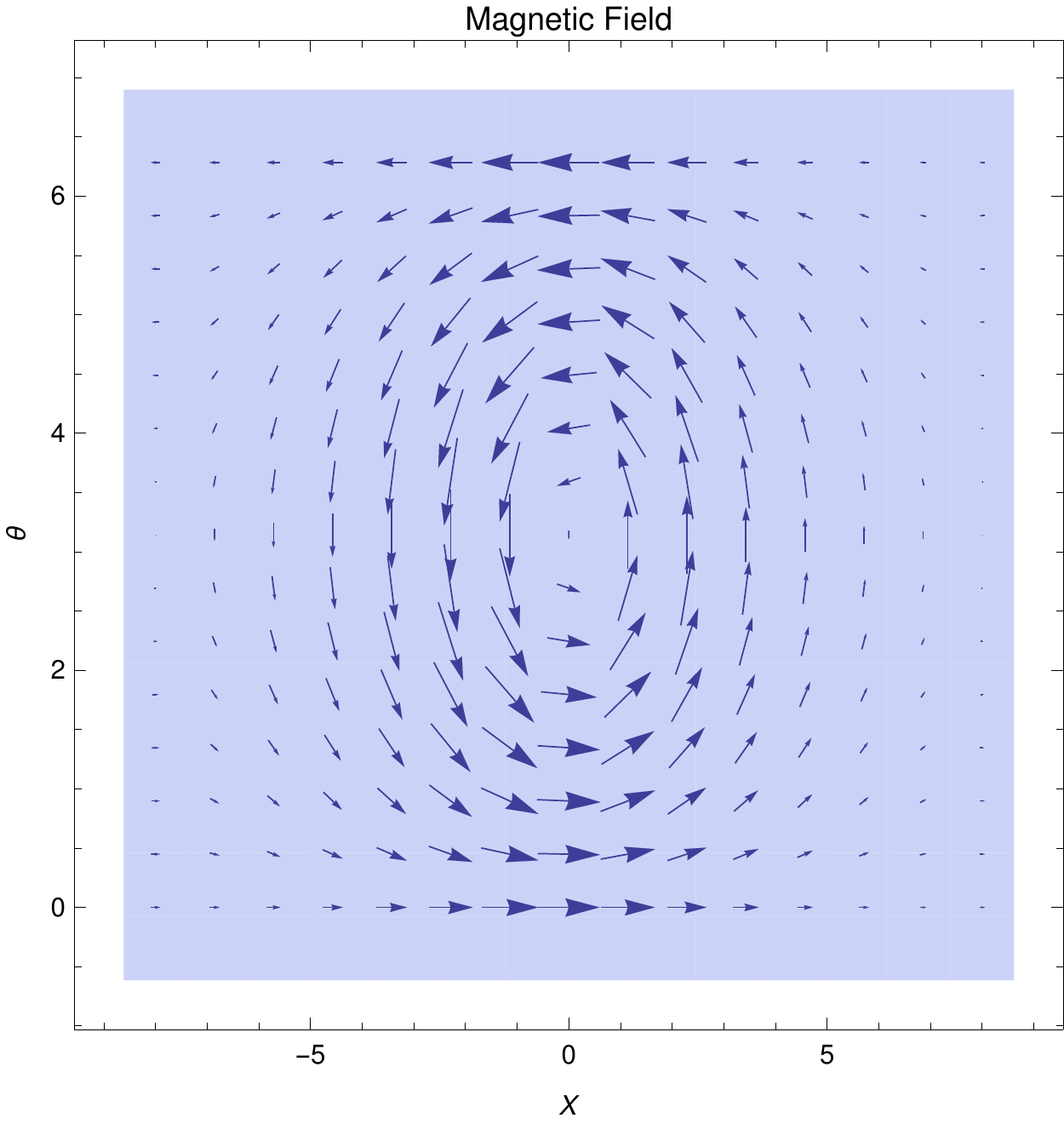}\ \ %
\includegraphics[scale=.4]{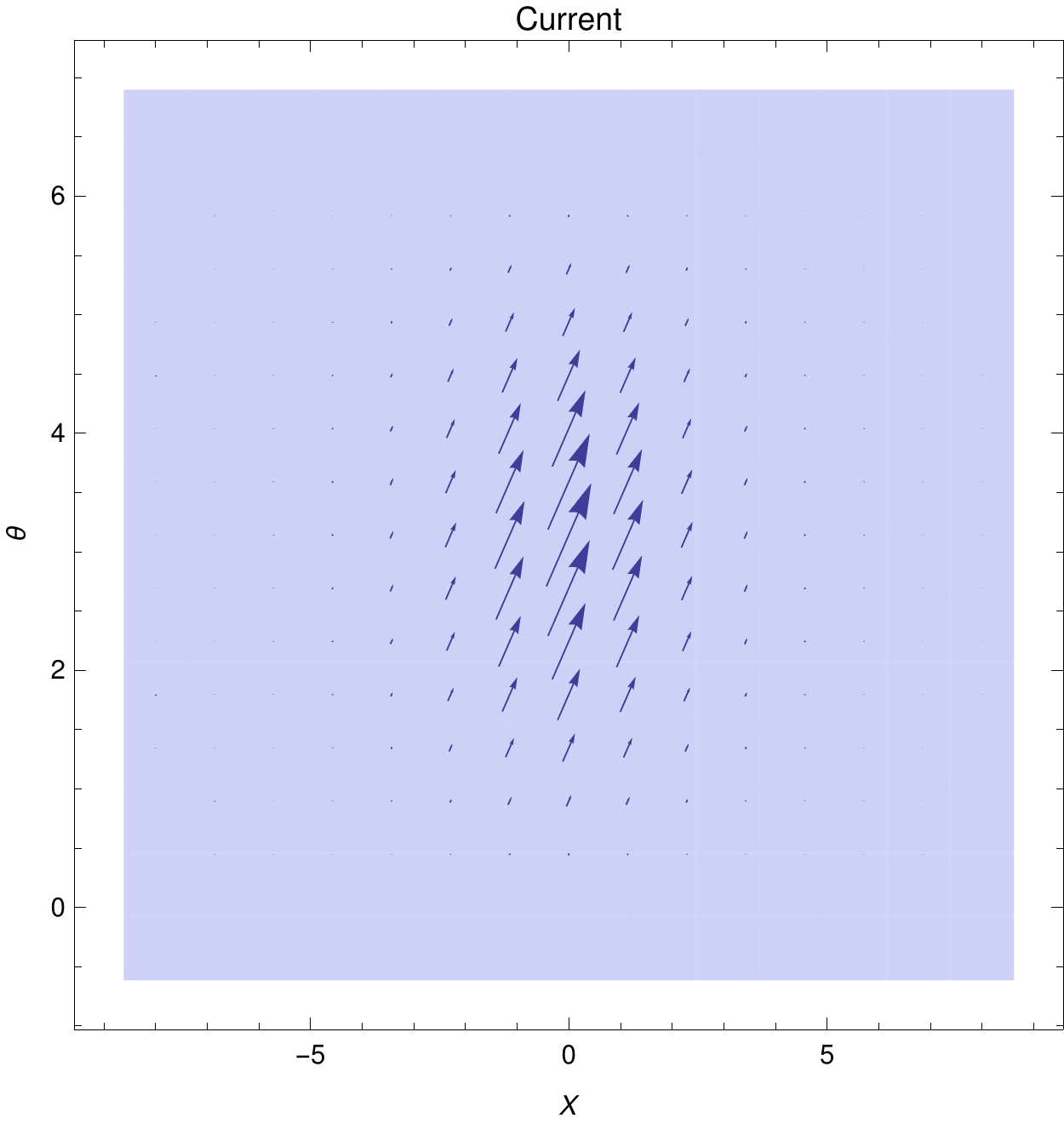}
\caption{From left to right, the vector density plots of the Electric field $%
E$, the magnetic field $B$ and of the current $J_\protect\mu$.}
\label{fig:EBJ}
\end{figure}
\noindent From the plots in Fig. \ref{fig:EBJ} one can see that $J_\mu\neq 0$
only at the position of the two peaks and tends to zero out. \noindent Here
we have imposed the following boundary conditions, 
\begin{gather*}
u(X, 2\pi)-u(X,0)=0\ ,\quad G(X, 2\pi)-G(X,0)=0\ .
\end{gather*}

\subsection{Why the tubes are superconductors?}

\subsubsection{Review of the Witten argument}

Before discussing the superconducting nature of the present solutions, we
will shortly review the results in \cite{wittenstrings} (which have been
considerably generalized in many subsequent works: see for instance \cite%
{subsequent1}, \cite{subsequent2}, \cite{subsequent3}, \cite{subsequent4}, 
\cite{subsequent5}, \cite{subsequent6}, and references therein).

The main motivation to introduce such topological objects is related, of
course, to the spectacular observable effects that such objects could have
(were they to exist: see the original reference \cite{wittenstrings} as well
as \cite{supstr1}, \cite{importance1}, \cite{importance2}, \cite{importance3}%
, \cite{importance4}, \cite{importance5}, \cite{importance6}, \cite%
{importance7}, \cite{new10}, \cite{new11}, \cite{new12}, \cite{new13}, \cite%
{new14}, \cite{new15}, \cite{new16}, \cite{new17}, \cite{new18}, \cite{new19}%
\ and references therein). The second motivation is related to the fact that
such remarkable objects can be constructed using quite reasonable
ingredients. Many of the examples available in the literature do not use
exclusively building blocks within the standard model\footnote{%
The stability properties of those have been considered, in \cite{Garaud:2008}%
, \cite{Forgacs:2009}, \cite{Hartmann:2012}. These ``twisted strings" were
found to be unstable. As rightfully remarked by the anonymous referee, one
should not expect adding further fields (considering the full standard
model) to stabilize these solutions.}. For instance, extra $U(1)$ gauge
potential as well as Higgs-like scalar fields are often important
ingredients while in \cite{subsequent1}, \cite{subsequent2}, \cite%
{subsequent3}, \cite{subsequent4}, \cite{subsequent5}, \cite{subsequent6},
supersymmetry plays a fundamental role.

The starting point of \cite{wittenstrings} is the following Lagrangian%
\footnote{%
We will change the notation of \cite{wittenstrings} slightly in order to
avoid confusions at later stages.} 
\begin{gather}
L_{\text{kin}} =-\frac{1}{4}\left( F^{2}+B^{2}\right) +\left\vert
D\sigma\right\vert ^{2}+\left\vert D\psi\right\vert ^{2}\ ,  \label{L1} \\
F_{\mu\nu} =\partial_{\mu}A_{\nu}-\partial_{\nu}A_{\mu}\ , \ \ B_{\mu\nu
}=\partial_{\mu}S_{\nu}-\partial_{\nu}S_{\mu}\ ,  \notag \\
D_{\mu}\sigma =\left( \partial_{\mu}+ieA_{\mu}\right) \sigma\ , \ D_{\mu
}\psi=\left( \partial_{\mu}+ieS_{\mu}\right) \psi\ .  \notag
\end{gather}

The above theory has an $U(1)\times U(1)$ gauge symmetry (the first
corresponding to $A_{\mu}$ and the second to $S_{\mu}$). In order for the
above theory to support superconducting strings, it is necessary to include
an interaction potential between the two Higgs fields $\sigma$ and $\psi$.
The choice of \cite{wittenstrings} was%
\begin{gather}
V\left( \sigma,\psi\right) =\frac{\lambda}{8}\left( \left\vert
\psi\right\vert ^{2}-\mu^{2}\right) ^{2}+\frac{\widetilde{\lambda}}{4}%
\left\vert \sigma\right\vert ^{4}+f\left\vert \sigma\right\vert
^{2}\left\vert \psi\right\vert ^{2}-m^{2}\left\vert \sigma\right\vert ^{2}\ ,
\label{L2} \\
S_{\text{tot}} =\int d^{4}x\sqrt{g}\left( L_{\text{kin}}+V\left(
\sigma,\psi\right) \right) \ .  \label{L3}
\end{gather}

The reasons behind this choice are the following: The first necessary
ingredient is the breakdown of the gauge symmetry corresponding to $S_{\mu }$
in order to ensure the existence of vortices. The Higgs field $\psi $ in the
core of the vortex field is usually assumed to only depend on the two
spatial coordinates (say, $r$ and $\theta $) transverse to the vortex axis
(which is along the $z$-axis). Then, one must require that, in the vacuum, $%
\left\langle \sigma \right\rangle =0$. At this point, with a clever choice
of the range of the parameters of the Higgs potential, one can achieve the
following situation. Despite the fact that the (minimization of the) kinetic
energy tends to suppress $\left\langle \sigma \right\rangle \neq 0$\ within
the core, if one chooses $m^{2}$ to be positive, the potential energy will
favor $\left\langle \sigma \right\rangle \neq 0$\ within the core. As it was
shown in \cite{wittenstrings}, this can indeed happen. In other words, there
is an open region in the parameter space in which $\left\langle \sigma
\right\rangle =0$ asymptotically but $\left\langle \sigma \right\rangle \neq
0$\ within the core of the vortex associated to $\psi $. This is a
fundamental technical step since the superconducting currents (to be
described in a moment) are sustained by the region in which $\left\langle
\sigma \right\rangle \neq 0$. If $\sigma _{0}\left( r,\theta \right) $
minimizes the energy of the string, then the superconducting current is
associated with the (slowly varying) phase $\Theta $ of $\sigma _{0}\left(
r,\theta \right) $. One can achieve this by introducing the dependence on $z$
and $t$ in $\sigma $ as follows:%
\begin{equation}
\sigma \left( r,\theta ,z,t\right) =\sigma _{0}\left( r,\theta \right) \exp 
\left[ i\Theta \left( z,t\right) \right] \ .  \label{wittsc}
\end{equation}%
The expression of the current is%
\begin{equation}
\overrightarrow{J}\approx 2e\sigma \left( \overrightarrow{\partial }\Theta +e%
\overrightarrow{A}\right) \ ,  \label{wittsc2}
\end{equation}%
which is made of two factors. The first factor $\left( \overrightarrow{%
\partial }\Theta +e\overrightarrow{A}\right) $\ is responsible for the
dynamics of the zero modes along the strings (associated to the phase $%
\Theta $). However, such a factor by itself would be ``useless" as it needs
to ``rely on" something. This ``something" is the factor $\sigma $. Thus,
first of all, the first factor needs $\sigma $ to be different from zero
somewhere. For superconducting strings, as it has been already emphasized, \
the spatial region where $\sigma \neq 0$ is tube-shaped. However, this is
not enough: the configurations in which $\sigma \neq 0$ within a tube-shaped
region must be stable, otherwise the current would decay\footnote{%
As it will be explained in the next subsection, the best option would be a
setting in which a suitable non-vanishing topological charge enforces $%
\sigma $, at the same time, to be different from zero in some spatial region
and to approach to zero outside. In this way, topology would ensure the
stability of the superconducting current.}. In the settings of \cite%
{wittenstrings}, \cite{subsequent1}, \cite{subsequent2}, \cite{subsequent3}, 
\cite{subsequent4}, \cite{subsequent5}, \cite{subsequent6}, the linear
stability of the configurations where $\sigma \neq 0$ within the string (as
well as $\sigma $ approaching to zero outside) were established by direct
methods (such as linear perturbation theory). Note also that the current
defined above cannot have arbitrarily large values since $\sigma $ has a
maximum value determined by the Higgs potential. Once the stability of such
tube-shaped regions (which are going to host the superconducting currents)
has been established, one can ask:

\textit{Under which circumstances the above current is superconducting?}

One needs a mechanism that keeps such current perpetually alive even in the
absence of an external gauge potential. At this point, topology comes into
play. Since $\Theta $ is only defined modulo $2\pi $, the integral over a
close loop of the above current will not vanish in general even when one
turns off the electromagnetic field. Indeed, one can build a topological
invariant associated to (the integral of) $\Theta $. Thus, such a current
cannot relax when the topological invariant associated to $\Theta $ is
non-zero.

\textbf{\textit{Basic ingredients}.} A very nice pedagogical construction
has been achieved in \cite{Shifman:2012vv}\ where, \textit{assuming that
symmetry breaking is always generated by suitable Higgs or scalar potentials}%
, the author reduced the basic ingredients to the skeleton. The author
required that bulk theory should have an unbroken global non-Abelian
symmetry (let us assume, just for concreteness, that such a group is $SU(2)$%
) allowing the existence of string-like configurations. Moreover, on such
string-like configurations, one has to break $SU(2)$ down to a subgroup $U(1)
$. As it has been already emphasized, such requirements are usually taken
care of by introducing suitable potentials for the scalars. In fact, the
Einstein-Maxwell-NLSM has all the above ingredients already ``built-in" and
there is no need to introduce any potential: the typical interactions among
the Maurer-Cartan forms associated to the Isospin degrees of freedom do the
job.

Secondly, it would be very nice to use topology not only to ensure the
persistent character of the current but also to guarantee the appearance of
regions where $\sigma \neq 0$ (such that $\sigma $ approaches zero outside).
Higher topological charges can be useful, as we will discuss in the
following subsection.

\subsubsection{The superconductivity of the tubes}

The main features of the above expressions in Eq. (\ref{currgaugedtubes})
for the current of these topologically non-trivial gauged crystals are the
following.

\textbf{1)} The current does not vanish even when the electromagnetic
potential vanishes ($u=0$).

\textbf{2)} Such a ``current left over" $J_{(0)\mu }$:%
\begin{equation}
J_{(0)\mu }=2K\sin ^{2}(\alpha )\sin ^{2}(q\theta )\partial _{\mu }\Phi \ ,
\label{left3}
\end{equation}%
(where $\Phi $ has been defined in Eq. (\ref{ansatz})) which survives even
when the Maxwell field is turned off, is maximal where the energy density is
maximal (namely, where $\sin ^{2}(\alpha )\sin ^{2}(q\theta )=1$, which
defines the positions of the peaks in the energy density as well as in the
topological density) and vanishes rapidly far from the peaks.

\textbf{3)} Such residual current $J_{(0)\mu }$ \textit{cannot be turned off
continuously}. This can be seen as follows. There are three ways to ``kill" $%
J_{(0)\mu }$. \textit{The first way} is to deform $\alpha $ to an integer
multiple of $\pi $ (but this is impossible as such a deformation would
change the topological charge). \textit{The second way} is to deform $%
q\theta $ to an integer multiple of $\pi $ (but also this deformation is
impossible due to the conservation of the topological charge). \textit{The
third way} is to deform $\Phi $ to a constant (but also this deformation
cannot be achieved). Note also that $\Phi $ is defined modulo $2\pi $ (as
the $SU(2)$ valued field $U$ depends on $\cos \Phi $ and $\sin \Phi $ rather
than on $\Phi $\ itself). This implies that the line integral of $\partial
_{\mu }\Phi $ along a closed contour does not necessarily vanish (as it
happens in the original Witten argument).

The above characteristics show that the above residual current is a
persistent current that cannot vanish as it is topologically protected.
This, by definition, implies that $J_{(0)\mu }$ defined in Eq. (\ref{left3})
is \textit{a superconducting current supported by the present gauged tubes}.

\section{The flat limit}

\label{secFlatLimit}

In this Section we discuss the flat limit $\kappa \rightarrow 0$ of our
solution describing a gauged gravitating soliton.  In this limit, 
the field equations in \eqref{Eqalpha}, \eqref{EqR} and %
\eqref{maxremaining} become 
\begin{align}
\alpha ^{\prime \prime }-\frac{q^{2}}{2}\sin (2\alpha )& \ = \  0\ ,
\label{flatALPHA} \\
\Delta G & \  = \ 0 \ ,  \\
R^{\prime \prime }& \ = \ 0\ , \\
\Delta u-2Ke^{-2R}(\omega _{s}-2u)\sin ^{2}{\alpha }\sin ^{2}{q\theta }&  \ = \ 0 \ .
\end{align}%
The solutions to the second and third equations are found to be 
\begin{equation}
G=0\ , \qquad R(X)=C_{2}X+C_{3}\ ,  \label{flatlimitR}
\end{equation}%
which corresponds to the flat limit of the solution in Eq. \eqref{R}. When 
$C_{2}$ does not vanish, the flat limit of the coordinate transformation to
the radial coordinate of a cylindrical system in Eq. \eqref{coordtrans} gives 
\begin{equation*}
r(X)=\frac{1}{-C_{2}}e^{-C_{2}X-C_{3}}\ ,
\end{equation*}%
so that the space-time metric becomes 
\begin{equation}
ds^{2}=-d\tilde{t}^{2}+d\tilde{z}^{2}+dr^{2}+C_{2}^{2}r^{2}d\theta ^{2}\ ,
\label{simpleconicaldefect}
\end{equation}%
where $\tilde{t}=\sqrt{2B_{0}\omega _{s}}t+\sqrt{B_{0}/2}z$, and $\tilde{z}=%
\sqrt{B_{0}/2}z$. To avoid a conical singularity, we should impose that 
\begin{align}
C_{2} = - 1 \ .
\end{align}
We have taken the negative value of $C_{2}$ to satisfy the inequality \eqref{condfinal}.
In this cylindrical coordinate system, the equations for $\alpha$ and $u$ become
\begin{align}
\Big( \frac{d^{2}}{dr^{2}} + \frac{1}{r} \frac{d}{dr} \Big) \alpha(r) 
- \frac{q^{2}}{2 r^{2}}\sin 2\alpha(r) = 0 \ , 
\label{flatalhpaeqn}
\\
\Big( \frac{\partial^{2}}{ \partial r^{2}} + \frac{1}{r} \frac{\partial}{ \partial r} 
+\frac{1}{r^{2}} \frac{\partial^{2}}{\partial \theta^{2}}\Big) u(r,\theta) 
- \frac{ 8 K (\omega _{s}-2u(r,\theta)) \sin ^{2}{q\theta } }
{\big( e^{|q|C_{3} + C_{1}} r^{|q|} + e^{-|q|C_{3} - C_{1}}  r^{-|q|}
 \big)^{2}} =  0 \ .
 \label{flatueqn1}
\end{align}
The constraint on the parameters \eqref{condfinal} restricts the range of the 
parameter $q$
\begin{align}
\frac{1}{1 - K \kappa} < |q| < \frac{1}{K \kappa} \ .
\label{qconst}
\end{align}
It should be noticed that $|q|=1$ is forbidden in the flat spactime limit if
$C_{2} \neq 0$.
The flat limit of the energy-density is found to be
\begin{align}
T_{\hat{0}\hat{0}} =  \frac{4 K \sin^{2}(q\theta)(2u-\omega_{s} )^{2}
+ K q^{2}/r^{2} }{\big( e^{|q|C_{3} + C_{1}} r^{|q|} + e^{-|q|C_{3} - C_{1}}  r^{-|q|}
 \big)^{2}}  +\frac{K}{2} \Big( %
(\partial_{r}u)^{2} +\frac{1}{r^{2}}(\partial_{\theta}u)^{2} \Big) \ ,
\label{flated}
\end{align}
where $\hat{\mathbf{e}}_{0} = \partial / \partial \tilde{t} $.
It follows from Eq. \eqref{flatueqn1} that the energy density \eqref{flated} of the 
gauged soliton decays rapidly, whereas the slow radial decay of the scalar field in the
Cohen-Kaplan solution results in the formation of a singularity at a certain distance
$r=r_{\textrm{max}}$ from the core of the cosmic string \cite{cohen-kaplan}. 
It should be noticed that $r_{\textrm{max}}$ is not the location of the maximum of 
the energy density (that is close to the string core) but a maximal radius in a non-singular 
part of spacetime, including the core of the Cohen-Kaplan string. \footnote{We would like 
to thank the anonymous referee for the constructive comments on this issue.}

Since we are seeking a regular solution for the topological solitons, 
it should be possible to expand $u(r, \theta)$ near the axis as
\begin{align}
u = r^{m} f(\theta) + O(r^{m+1}) \ ,
\end{align} 
for some constant $m$ and some smooth function $f(\theta)$. 
It follows from Eq. \eqref{flated} that the energy-density of the system 
diverges at $r=0$ unless $m \geq 1$. 
Near the axis, Eq. \eqref{flatueqn1} at leading order is found to be
\begin{align}
\Big( m^{2} + \partial_{\theta}^{2}  \Big) f(\theta)
- 8 K e^{2|q| C_{3} + 2 C_{1}} \sin^{2}(q\theta) \  r^{2|q|+2} \
\Big( \frac{\omega_{s}}{r^{m}} - 2 f(\theta)  \Big) = 0 \ .
\end{align}
This equation admits a solution for an arbitrarily small $r$ only  when
$q=0$, which is forbidden by \eqref{qconst}. For this reason, the flat limit 
of our solution for the gauged soliton is singular if we assume that $C_{2} \neq 0$.

Therefore, the genuine flat limit of the gravitating superconducting tubes
described in the previous sections is obtained when the integration constant 
$C_{2}$ in Eq. (\ref{flatlimitR}) vanishes. In this way we recover the
configurations of topologically non-trivial gauged solitons within the
space-time metric given by 
\begin{equation}
ds^{2}=-dt^{2}+dX^{2}+L^{2}\left( d\theta ^{2}+d\phi ^{2}\right) \ ,
\label{Minkowski}
\end{equation}%
which was studied in \cite{crystal2} (but with a difference, which we will see immediately). 

In order to shed further light on the subtleties of the flat limit, it is
worth to remind that the analytic solutions in \cite{crystal2} represent
gauged superconducting tubes in a box with finite volume described by the
flat metric%
\begin{equation}
ds^{2}=-dt^{2}+L^{2}\left( dr^{2}+d\theta ^{2}+d\phi ^{2}\right) \ ,
\label{minkowski2}
\end{equation}%
in which the three coordinates ($r$, $\theta $ and $\phi $) have finite
range in order to describe a situation with a finite amount of Baryonic
charge within a finite spatial volume. Thus, at least locally, one can
identify the coordinate $X$ with the coordinate $r$ in \cite{crystal2}. The
ansatz in \cite{crystal2} gives rise to Eq. (\ref{flatALPHA}) for the
profile $\alpha $ which can be easily reduced to the following first order
equation 
\begin{equation}
(\alpha ^{\prime })^{2}-q^{2}\sin ^{2}{\alpha }\ =E_{0}\ ,
\label{flatALPHA2}
\end{equation}
where $E_{0}$ is an integration constant. As explained in details in \cite%
{crystal2}, the role of the integration constant $E_{0}$ is very important
since it allows to ``squeeze" an arbitrary amount of Baryonic charge within
the finite flat box\footnote{%
In \cite{crystal2}, $E_{0}$
can be fixed in such a way that the peaks of the energy density are periodic
in $r$ so that the system manifests a crystalline structure. Indeed, in the
flat case the Baryonic charge is proportional to the number of peaks (labelled by $n$ in Eq. \eqref{E0flat}) of the
energy density in the $r$ direction.} and it is fixed through the relation
\begin{equation} \label{E0flat}
 n\int_0^{\pi} \frac{1}{\eta(\alpha,E_0)}=2\pi \ , \qquad \eta(\alpha,E_0)=\pm\biggl[ 2\biggl( E_0-\frac{q^2}{4}\cos(2\alpha)+\frac{4m^2}{K}\cos(\alpha)  \biggl) \   \biggl]^{\frac{1}{2}} \ . 
\end{equation}
It is also important to remember that in the flat case studied in \cite{crystal2} the Maxwell equations are reduced to a single equation in the form
\begin{equation} \label{uflat}
 \Delta \Psi + V \Psi = 0 \ , \qquad V= 4L^2 K \sin^2(\alpha) \sin^2(q\theta) \ , 
\end{equation}
with $\Psi=\frac{2L}{p}u-1$.

On the other hand, there is a subtle fingerprint of the Einstein equations
which does not disappear in the flat limit. The reason is that the profile $%
\alpha $ corresponding to the (flat limit of the) gravitating hadronic tubes
constructed in the previous sections corresponds to a very precise choice of
the integration constant $E_{0}$ in Eq. (\ref{flatALPHA2}). It is easy to
see, comparing Eqs. (\ref{Eqalpha}) and (\ref{Eqalpha0}) with the
equation for $\alpha $ in the flat case in \cite{crystal2}, that the flat limit of the present
gravitating superconducting tubes corresponds to the case $E_{0}$=0, namely  
\begin{equation*}
(\alpha ^{\prime })^{2}-q^{2}\sin ^{2}{\alpha }\ =E_{0}=0\ .
\end{equation*}%
The fact that $E_{0}$ vanishes has two important consequences. Firstly, it admits only one peak of the energy density along the direction
of $X$ as could be expected. Secondly, the solution for $\alpha$ necessarily describes a kink from $-\infty $ to $+\infty $%
. The finite range of the coordinate $\theta $ is consistent with the fact
that, when the flat superconducting tubes in \cite{crystal2} are promoted to
gravitating superconducting tubes as described in this manuscript, the
coordinate $\theta $  becomes an angular coordinate. Hence, the flat limit
of the gravitating solitons turn out to be a single superconducting tube in
a box that has been extended to infinity in the $X$-direction.

The analysis of the flat solutions in \cite{crystal2} also clarifies why the
gravitating version of the flat gauged solitons presented
in this manuscript is not axi-symmetric. The reason is already manifest in
the (flat limit of the) energy-density:%
\begin{equation*}
T_{\hat{0}\hat{0}}=K\Big[q^{2}\sech^{2}(qX+C_{1})+\sin ^{2}(q\theta )\sech%
^{2}(qX+C_{1})\Big]\ ,
\end{equation*}%
(where we have used the freedom to scale some integration constants to 1).
Clearly, the above expression (which is the energy density of the flat limit
of the present gravitating hadronic tubes) is not axi-symmetric. The obvious
reason is the explicit dependence of the energy-density on $\theta $ (which,
in the gravitating case, plays the role of angular coordinate). Thus, in a
sense, it is natural that the gravitating version of the flat solitons in 
\cite{crystal2} should not be axi-symmetric. We see also that $T_{\hat{0}\hat{0}}$ decays rapidly in $X$ as expected.
Unlike what happens with the Cohen-Kaplan solution \cite{bookVS}, our configurations have a maximum at a finite distance while in Cohen-Kaplan this is a delta, being the difference exponentially small. In fact, we can say that our solution approaches asymptotically to the Cohen-Kaplan solution.

The above arguments show that our solution exists for an arbitrarily small $%
\kappa $. As can be seen from the curvature invariants given in Section \ref%
{regularity}, the gravitational effects on the curvature would be small in
the weak-field limit. It should be emphasized that, however, the most
spectacular consequence of the gravitating tube is the large deviation of
the worldlines of physical particles (such as photons and electrons) from
the geodesics. In particular, photons will not follow null geodesics due to
the effective mass term arising due to the minimal coupling with the
NLSM. On the other hand, electrons will not follow
time-like geodesics since they will be accelerated by the Lorentz force.
Both effects will be especially strong close to the local maxima of the
energy-density. Consequently, the observer at infinity will detect a sudden
and large deviations of these particles from the geodesic motion in an
almost flat space-time.

\section{Conclusions}

The first example of analytic and curvature singularity free cosmic tube
solutions for the Einstein $SU(2)$-NLSM has been found. The metric at large
distance from the axis looks similar to a boosted cosmic string. The matter
distribution has no sharp boundary and the curvature is concentrated at a
finite distance from the axis. The angular defect of the solutions depends
on the distance from the axis and the parameters of the solutions can be
chosen in such a way that it vanishes near the axis but not at large
distance. These properties make the solution similar to a global string but
with the fundamental difference that while the global string has a curvature
singularity at a finite distance from the axis whereas the new solutions are
singularity free. \newline
Due to the non-Abelian symmetry group, the most natural way to impose a
periodicity condition on the $SU(2)$ field is up to an inner space rotation.
This very natural boundary condition allows the solution to carry
non-trivial topological charge.\newline
One of the most remarkable aspects of these solutions is that they can also
be promoted to solutions of the Einstein-Maxwell-NLSM theory, i.e. the $SU(2)
$ field is minimally coupled to both the $U(1)$ field and gravity, and
without neglecting the corresponding Maxwell equations ``sourced" by the
currents arising from the NLSM. The gauged solutions are characterized by
the fact that they can carry a persistent current even when the Maxwell
field is zero, which means that they are superconducting. Moreover, the
superconducting current is also topologically protected.\newline
It is worth pointing out that in cosmology one of the most important
observational consequences of the existence of cosmic strings is
gravitational lensing. The overwhelming majority of papers dealing with
gravitational lensing assume axi-symmetry of the cosmic string. The analytic
solution found here however is not axi-symmetric and therefore the geodesic
equation becomes non-trivial. An interesting feature of the solution found
here is that it is highly repulsive in the core of the tube, the matter
distribution has no sharp boundary and therefore spreads to infinity, the
deflection angle depends on the initial distance from the source. It is
reasonable to suppose that these non-trivial features should have interesting
observational consequences and will be an object of study in further
investigations.

\acknowledgments

We thank the anonymous referee for his/her detailed review and the suggestion 
to include in the manuscript the analysis of the flat limit, 
which helps to clarify several relevant aspects of the solutions presented here.
F. C. and A. G. have been funded by Fondecyt Grants 1200022 and 1200293. M.
L. and A. V. are funded by FONDECYT post-doctoral grant 3190873 and 3200884.
The Centro de Estudios Cient\'ificos (CECs) is funded by the Chilean
Government through the Centers of Excellence Base Financing Program of
Conicyt. S. H. O. is supported by the National Research Foundation of Korea
funded by the Ministry of Education of Korea (Grant 2018-R1D1A1B0-7048945,
2017- R1A2B4010738).

\appendix

\section{How the ansatz solves the system}

\label{App1}

In this first Appendix, we provide all the technical details behind the
ansatz that we have used in the main text to solve the system of coupled
field equations corresponding to the Einstein-Maxwell-NLSM together with the
technical steps needed to derive the field equations.

As it has been already mentioned in the main text, the action of the NLSM
and the field equations read, respectively%
\begin{gather}
I_{\text{NLSM}} = \int \sqrt{-g}\frac{K}{4}\text{Tr}\left[ L_{\mu }L^{\mu }%
\right] d^{4}x\ ,  \label{general0} \\
\nabla ^{\mu }\left( L_{\mu }\right) = 0\ ,  \label{general0.1}
\end{gather}%
where%
\begin{equation*}
L_{\mu }=U^{-1}\partial_{\mu }U\ ,\ U(x) \in SU(2)\ .
\end{equation*}%
In order to achieve a better understanding of the present framework and of
why our ansatz works it is convenient to express the above action explicitly
in terms of the most general parametrization of $SU(2)$:%
\begin{gather}
U(x^{\mu }) = \left( \cos \alpha \right) \mathbf{1}+\left( \sin \alpha
\right) n_{j}t^{j}\ ,\ t^{j}=i\sigma ^{j}\ ,  \label{general1} \\
n_{1} = \sin \Theta \cos \Phi \ ,\ n_{2}=\sin \Theta \sin \Phi \ ,\
n_{3}=\cos \Theta \ .  \notag
\end{gather}
Here the three $SU(2)$ Pionic degrees of freedom are encoded in the three
scalar functions $\alpha =\alpha (x^{\mu })$, $\Theta =\Theta (x^{\mu })$
and $\Phi =\Phi (x^{\mu })$. Consequently, the action of the NLSM in Eq. (%
\ref{general0}) can be expressed explicitly in terms of the three scalar
functions $\alpha$, $\Theta$ and $\Phi$. A direct computation shows that the
result is the following action (which, of course, is completely equivalent
to the one in Eq. (\ref{general0})): 
\begin{equation}
I_{\text{NLSM}}=\int \sqrt{-g}\frac{K}{2}(\nabla _{\mu }\alpha \nabla ^{\mu
}\alpha +\sin ^{2}\alpha \nabla _{\mu }\Theta \nabla ^{\mu }\Theta +\sin
^{2}\alpha \sin ^{2}\Theta \nabla _{\mu }\Phi \nabla ^{\mu }\Phi )d^{4}x\ .
\label{general2}
\end{equation}%
It is very convenient also to express the field equations in Eq. (\ref%
{general0.1}) in terms of the three scalar degrees of freedom. A direct
computation shows that, when varying the action in Eq. (\ref{general2}) with
respect to $\alpha $, $\Theta $ and $\Phi $, the following field equations
are obtained 
\begin{gather}
\Box {\alpha }-\sin \alpha \cos \alpha (\nabla _{\mu }\Theta \nabla ^{\mu
}\Theta +\sin ^{2}\Theta \nabla _{\mu }\Phi \nabla ^{\mu }\Phi )=0\ ,
\label{Appeqs3} \\
\sin ^{2}\alpha \Box \Theta +2\sin \alpha \cos \alpha \nabla _{\mu }\alpha
\nabla ^{\mu }\Theta -\sin ^{2}\alpha \sin \Theta \cos \Theta \nabla _{\mu
}\Phi \nabla ^{\mu }\Phi =0\ ,  \label{appeqs3.1} \\
\sin ^{2}\alpha \sin ^{2}\Theta \Box \Phi +2\sin \alpha \cos \alpha \sin
^{2}\Theta \nabla _{\mu }\alpha \nabla ^{\mu }\Phi +2\sin ^{2}\alpha \sin
\Theta \cos \Theta \nabla _{\mu }\Theta \nabla ^{\mu }\Phi =0\ .
\label{appeqs3.2}
\end{gather}%
It is important to emphasize that the field equations in Eqs. (\ref{Appeqs3}%
), (\ref{appeqs3.1}) and (\ref{appeqs3.2}) are completely equivalent to the
field equations written in the more usual matrix form in Eq. (\ref%
{general0.1}). One can easily check that if $\alpha $, $\Theta $ and $\Phi $
satisfy Eqs. (\ref{Appeqs3}), (\ref{appeqs3.1}) and (\ref{appeqs3.2}) then
Eq. (\ref{general0.1}) are satisfied as well (and viceversa). The next piece
of information needed to build the ansatz is the topological charge density 
\begin{equation*}
\rho _{\text{B}}=12(\sin ^{2}{\alpha }\sin ^{2}{\Theta })d\alpha \wedge
d\Theta \wedge d\Phi \ .
\end{equation*}%
As we want to consider only topologically non-trivial configurations, we
must demand that 
\begin{equation}
d\alpha \wedge d\Theta \wedge d\Phi \neq 0\ .  \label{good3}
\end{equation}%
Now, the problem is to find a good ansatz which respects the above condition
and simplify as much as possible the field equations. A close look at Eqs. (%
\ref{Appeqs3}), (\ref{appeqs3.1}) and (\ref{appeqs3.2}) reveals that a good
set of conditions is 
\begin{equation}
\nabla _{\mu }\Phi \nabla ^{\mu }\alpha =\nabla _{\mu }\alpha \nabla ^{\mu
}\Theta =\nabla _{\mu }\Phi \nabla ^{\mu }\Phi =\nabla _{\mu }\Theta \nabla
^{\mu }\Phi =0\ .  \label{Appconds}
\end{equation}%
Moreover, one should not forget the Derrick's no-go theorem which prevents
the existence of static solitonic solutions of the NLSM: thus, the ansatz
should depend on time but in such a way to have a stationary energy momentum
tensor (as expected from a superconductive tube). A choice that satisfies
Eq. \eqref{Appconds} and, at the same time, circumvent Derrick's scale
argument is 
\begin{equation}
\alpha =\alpha (X)\ ,\ \ \ \Theta =q\theta \ ,\ \ \ \Phi =\omega _{s}t+z\ ,
\label{good1}
\end{equation}%
where the space-time metric\footnote{%
In this Appendix we will always refer to that class of space-time metrics.}
and the coordinates $r$, $\theta $, $t$ and $z$ have been defined in the
main text in Eq. (\ref{WLP}). Some other useful identities satisfied by the
above ansatz are 
\begin{equation}
\Box \Theta =\Box \Phi =0\ ,  \label{good2}
\end{equation}%
together with Eq. \eqref{good3}.

Such identities greatly simplify the field equations. A direct computation
reveals that the three NLSM field equations Eqs. (\ref{Appeqs3}), (\ref%
{appeqs3.1}) and (\ref{appeqs3.2}) are reduced to the following ODE for $%
\alpha $:%
\begin{equation}
\alpha ^{\prime \prime }=\frac{q^{2}}{2}\sin \left( 2\alpha \right) \ ,
\label{goodalpha}
\end{equation}
keeping alive, at the same time, the topological charge.

Remarkably enough, these very intriguing properties of the ansatz are not
destroyed by the inclusion of the minimal coupling with Maxwell field. The
coupling of the NLSM with the Maxwell theory is introduced replacing, in the
action in Eq. (\ref{general0}), the partial derivatives acting on the $SU(2)$%
-valued scalar field $U$ with the following covariant derivative 
\begin{equation}
\partial_{\mu }U\rightarrow D_{\mu }U=\partial_{\mu }U+A_{\mu }[t_{3},U]\ .
\label{covdev1}
\end{equation}%
A straightforward computation shows that the above replacement in Eq. (\ref%
{covdev1}) in the Lagrangian in Eq. (\ref{general0}) is completely
equivalent to the replacement here below (in terms of $\alpha $, $\Theta $\
and $\Phi $)%
\begin{gather}
\partial _{\mu }\alpha \rightarrow \partial _{\mu }\alpha \ ,
\label{covdev2} \\
\partial _{\mu }\Theta \rightarrow \partial _{\mu }\Theta \ ,  \notag \\
\partial _{\mu }\Phi \rightarrow D_{\mu }\Phi =\partial _{\mu }\Phi -2A_{\mu
}\Phi \ ,  \notag
\end{gather}%
in the action in Eq. (\ref{general2}). It is worth to emphasize that $D_{\mu
}\Phi $ determines the \textquotedblleft direction\textquotedblright \ of
the electromagnetic current (as it will be discussed here below). Thus, in
this case the field equations for the gauged NLSM are%
\begin{equation}
D^{\mu }\left( U^{-1}D_{\mu }U\right) =0\ ,  \label{gaugedfe}
\end{equation}%
where, for notational simplicity, we will still define as%
\begin{equation*}
U^{-1}D_{\mu }U=L_{\mu }\ .
\end{equation*}
Obviously, when the derivative is replaced with the Maxwell covariant
derivative (as defined in Eq. (\ref{covdev1}) or, equivalently, in Eq. (\ref%
{covdev2})), in the field equations of the gauged NLSM many new terms will
appear, coupling the $SU(2)$ degrees of freedom with the $U(1)$ gauge
potential. Thus, one may ask: \textit{Which is the best choice of the ansatz
for the gauge potential} $A_{\mu }$ \textit{that keeps as much as possible
the very nice properties of the ansatz of the} $SU(2)-$\textit{valued scalar
field in Eqs. \eqref{good3}, \eqref{Appconds} and \eqref{good2}, and allowed
(as explained in the main text) the complete analytic solutions in the
Einstein-NLSM case}?

In order to achieve this goal, it is enough to demand 
\begin{gather}
\nabla ^{\mu }A_{\mu }=0\ ,\ A_{\mu }A^{\mu }=0\ ,\ A_{\mu }\nabla ^{\mu
}\Phi =0\ ,  \label{good4} \\
A_{\mu }\nabla ^{\mu }\alpha =0\ ,\ A_{\mu }\nabla ^{\mu }\Theta =0\ .
\label{good5}
\end{gather}%
The above conditions determine that the Maxwell potential $A_{\mu }$ must be
of the form 
\begin{equation}
A_{\mu }=\biggl(u\left( X,\theta \right) ,0,0,\frac{1}{\omega _{s}}u\left(
X,\theta \right) \biggl)\ .  \label{AAmugrav}
\end{equation}%
From the above, the explicit expressions of the components $L_{\mu }$ and $%
J_{\mu }$ can be evaluated directly as 
\begin{equation*}
{\small L_{t}=(\omega _{s}-2u)%
\begin{pmatrix}
\hspace{-1in}i\sin ^{2}\alpha \sin ^{2}(q\theta ) & \hspace{-1in}%
e^{-i(z+t\omega _{s})}\sin \alpha \sin (q\theta )(\cos \alpha -i\sin \alpha
\cos (q\theta ) \\ 
-e^{i(z+t\omega _{s})}\sin \alpha \sin (q\theta )(\cos \alpha +i\sin \alpha
\cos (q\theta ) & -i\sin ^{2}\alpha \sin ^{2}(q\theta )%
\end{pmatrix}%
\ , }
\end{equation*}%
\begin{equation*}
{\small L_{X}=%
\begin{pmatrix}
i\cos (q\theta )\alpha ^{\prime } & e^{i(z+t\omega _{s})}\sin (q\theta
)\alpha ^{\prime } \\ 
ie^{i(z+t\omega _{s})}\sin (q\theta )\alpha ^{\prime } & -i\cos (q\theta
)\alpha ^{\prime }%
\end{pmatrix}%
\ ,}
\end{equation*}%
\begin{equation*}
{\small L_{\theta }=%
\begin{pmatrix}
-iq\sin \alpha \cos \alpha \sin (q\theta ) & qe^{-i(z+\omega _{s}t)}\sin
\alpha (\sin \alpha +i\cos \alpha \cos (q\theta )) \\ 
-qe^{i(z+\omega _{s}t)}\sin \alpha (\sin \alpha -i\cos \alpha \cos (q\theta
)) & iq\sin \alpha \cos \alpha \sin (q\theta )%
\end{pmatrix}%
\ ,}
\end{equation*}%
\begin{equation*}
L_{z}=\frac{1}{\omega _{s}}L_{t}\ ,
\end{equation*}%
\begin{equation*}
J_{t}=2K\sin ^{2}(\alpha )\sin ^{2}(q\theta )(\omega _{s}-2u) \ , \quad
J_{X}=J_{\theta }=0 \ ,\quad J_{z}=\frac{1}{\omega_s}J_{t} \ ,
\end{equation*}%
\begin{equation}
\Leftrightarrow J_{\mu }=2K\sin ^{2}(\alpha )\sin ^{2}(q\theta )\left[
\partial _{\mu }\Phi -2A_{\mu }\Phi \right] \ .  \label{goodcurr1}
\end{equation}%
From the expressions for $L_{\mu }$ one can see that, despite the explicit
presence of $A_{\mu }$ in the $U(1)$-covariant derivative, the three field
equations of the gauged NLSM still reduce to Eq. (\ref{goodalpha}). The
reason is that all the potential terms which, in principle, could couple the 
$SU(2)$-valued scalar field $U$ with $A_{\mu } $ in the field equations in
Eq. (\ref{gaugedfe}) actually vanish due to the identities in Eqs. (\ref%
{good2}), (\ref{good4}) and (\ref{good5}) satisfied by the choice of our
ansatz (that is why we have chosen the ansatz in that way). Moreover, due to
the presence of a quadratic term in $A_{\mu }$ in the action in Eqs. (\ref%
{general0}) and (\ref{general2}), which couple $A_{\mu }$ with the $SU(2)$%
-valued scalar field $U$ (as it happens in the Ginzburg-Landau description
of superconductors), even when $A_{\mu }=0$ the current does not vanish (as
it is clear from Eq. (\ref{goodcurr1})). Such a residual current (which
survives even in the $A_{\mu }=0$\ limit) cannot be deformed continuously to
zero: the reason is that the only way to ``kill'' would also kill the
topological charge but, as it is well known, there is no continuous
transformation that can change the topological charge. Consequently, the
current in Eq. (\ref{goodcurr1}) is a persistent current. \noindent One can
easily verify that, thanks to the properties in Eqs. (\ref{good4}) and (\ref%
{good5}) satisfied by the $A_{\mu }$ in Eq. (\ref{AAmugrav}), the four
Maxwell equations with the current in Eq. (\ref{goodcurr1}) are reduced to
the single PDE in Eq. \eqref{maxremaining} of the main text. Finally,
evaluating the metric in Eq. \eqref{WLP} together with the ansatz for the
scalar field $U$ and the Maxwell potential $A_\mu$ in Eqs. \eqref{good1} and %
\eqref{AAmugrav} in the Einstein equations, one gets the equations for the
metric functions $R$ and $G$ defined in Eqs. \eqref{EqG} and \eqref{KK}.

\section{About the stability in the flat limit}

\label{App2}

A remark on the stability of the superconducting tubes in flat space-time constructed in \cite{crystal2} (which in the case $E_0=0$ is reduced to the flat limit of the gravitating tube) is in order.

\subsection{Perturbations on the profile}

In many situations, when the hedgehog property holds (so that the field
equations reduce to a single equation for the profile) the simplest
non-trivial perturbations are those of the profile which keep the structure
of the hedgehog ansatz (see \cite{shifman1}, \cite{shifman2} and references
therein). One of the key technical results of the present work is that such
a property holds even when the minimal coupling with GR is considered. In
particular, it holds in the flat case (as shown in \cite{crystal1}, \cite%
{crystal2}).

In the present case the simplest non-trivial perturbations are of the
following form:%
\begin{equation}
\alpha \rightarrow \alpha +\varepsilon \xi \left( X\right) \ ,\ \ \
0<\varepsilon \ll 1\ ,  \label{pert}
\end{equation}%
which do not change the Isospin degrees of freedom associated with the
functions $\Theta $\ and $\Phi $. The NLSM equations at first order in $%
\varepsilon$ are reduced to 
\begin{align*}
(\alpha^{\prime \prime }-\frac{q^2}{2}\sin(2\alpha))+(\xi^{\prime \prime
}-q^2\cos(2\alpha)\xi)\varepsilon = 0 \ .
\end{align*}
We can see that, imposing the on-shell relations we get to the following
equation for $\xi(X)$: 
\begin{align*}
\xi^{\prime \prime }-q^2\cos(2\alpha)\xi = 0 \ ,
\end{align*}
so that the system always has the following zero-mode: $\xi \left( X \right)
=\partial _{X}\alpha _{0}\left( X \right) $, with $\alpha_0$ a solution of
the NLSM equation. Due to the fact that the integration constant $E_{0}$ in Eqs. \eqref{flatALPHA2} and \eqref{E0flat} can always be
chosen in such a way that $\partial _{X}\alpha _{0}\left( X\right) $ never
vanishes (see \cite{crystal2}), the zero-mode $\xi \left( X\right) $ has no node so that it must
be the perturbation with lowest energy. Thus, the present solutions are
stable under the above potentially dangerous perturbations. This is a very
non-trivial test.

Now we discuss the relation between the stability and the existence of a
topological charge. The on-shell Hamiltonian density $\mathcal{H}$ obtained
evaluating the ansatz in Eqs. \eqref{general1} and \eqref{good1} in Eq. %
\eqref{general0.1} turns out to be 
\begin{align*}
\mathcal{H}_{\text{on-shell}} = \frac{K}{2L^2}(\alpha^{\prime
2}+q^2\sin^2(\alpha))\ .
\end{align*}
Consequently, the on-shell energy $E$ of the system can be written as%
\begin{equation}
E_{\text{on-shell}} = \pi^{2} K L \int \left[(\alpha^{\prime 2}\mp
2q\sin(\alpha)\alpha^{\prime }\right] dX\ ,
\end{equation}%
which turns out to be a positive definite term plus a total derivative. The
integral of this total derivative defines a new topological charge $Q(n)$,
namely 
\begin{equation*}
Q(n)=2q\int_{0}^{2\pi }\left( \sin \alpha \right) \alpha ^{\prime }dX=2q\cos{%
\alpha(0)} \left[ 1-\left( -1\right) ^{n}\right] \ ,
\end{equation*}%
which depends on $n$. It is worth to note that the above topological charge $%
Q(n)$ is different from the Baryonic charge as the Baryonic charge is
non-vanishing for any integer $n$ appearing in the boundary conditions (see \cite{crystal2}) while the above topological charge is
non-vanishing only when $n$ is odd. The corresponding BPS bound for the
on-shell energy is, 
\begin{equation}
E_{\text{on-shell}} \geq \pi^{2} K L \left\vert Q(n)\right\vert \ .
\label{bpstyle3}
\end{equation}%
One can see that there exist configurations which saturate the bound: such
configurations satisfy%
\begin{equation}
\alpha ^{\prime }\pm q\sin \alpha =0\ .  \label{bpstyle4}
\end{equation}
Interestingly enough, also in this case the saturation of the above BPS
bound does imply the general field equations. Thus, the present analytic
solutions have two topological labels: the Baryonic charge $\left\vert w_{%
\text{B}}\right\vert =\left\vert np\right\vert $ and $Q(n)$. Obviously, the
configurations which saturate the above bound are stable. This happens
precisely when the integration constant $E_{0}$ in Eqs. (\ref%
{flatALPHA2}) and \eqref{E0flat} is such that the solution for $\alpha$ is also compatible with the Einstein
equations (see Eq. (\ref{EqR})). The present stability argument also extends
to the gravitating configurations constructed in the main text which have $%
\left\vert n\right\vert =1$. 

\subsection{Electromagnetic perturbations}

Another very useful approach to study the stability is to consider only
electromagnetic perturbations. For simplicity (see Eqs. \eqref{maxwellNLSM}, \eqref{current} and \eqref{uflat}), let us consider the following
type of electromagnetic perturbations: 
\begin{align*}
(u,0,0,-Lu) \rightarrow (u+\varepsilon c_1,0,0,-Lu+ \varepsilon c_2) \ ,
\quad c_i=c_i(t,X,\theta,\phi) \ , \quad \varepsilon\ll 1 \ .
\end{align*}
We will consider that only the electromagnetic field is affected but not the
solitons. This is a good approximation in the 't Hooft limit, since in the
semiclassic interaction photon-Baryon, the Baryon is essentially unaffected
(because the photon has zero mass) \cite{WeigelNotes}. Electromagnetic
perturbations, therefore, perceive the solitons as an effective medium and
this, in practical terms, greatly simplifies the stability analysis as we
will see below.

At first order in the parameter $\varepsilon$ the Maxwell equations in Eqs. \eqref{maxwellNLSM} and \eqref{current} become 
\begin{gather*}
\partial_\theta (\partial_\phi c_2 -L^2 \partial_t c_1) = 0 \ , \\
\partial_X (\partial_\phi c_2 -L^2 \partial_t c_1) = 0 \ , \\
\partial_\phi^2 c_1 + \partial_\theta^2 c_1 +\partial_X^2 c_1 -\partial_t
\partial_\phi c_2 +4eKL^2\sin^2(\alpha)\sin^2(q\theta) c_1 = 0 \ , \\
\partial_\theta^2 c_2 + \partial_X^2 c_2 -L^2 \partial_t^2 c_2 +
L^2\partial_t \partial_\phi c_1 +4eKL^2\sin^2(\alpha)\sin^2(q\theta) c_2 = 0
\ .
\end{gather*}
It is worth to emphasize the following fact. To test linear stability, one
should check the (absence of) growing modes in time. Therefore, the two
perturbations $c_1$ and $c_2$ must depend on time. On the other hand, due to
the above equations, it is clear that if $c_1$ and $c_2$ depend on time then
they also depend on the coordinate $\phi$. Hence 
\begin{align}
\partial_{\phi} c_{1} \neq 0 \ , \qquad \partial_{\phi} c_{2} \neq 0 \ .
\end{align}
Now, assuming for simplicity that 
\begin{align*}
\partial_\phi c_2 = L^2 \partial_t c_1 \ ,
\end{align*}
the Maxwell equations at first order in $\varepsilon$ are reduced to 
\begin{align*}
\Box c_i + V c_i = 0 \ , \qquad \Box\equiv -L^2 \partial_t^2+\partial_X^2
+\partial_\theta^2 + \partial_\phi^2 \ ,
\end{align*}
with $V$ defined in Eq. \eqref{uflat} and $c_i=\{c_1,c_2 \}$.

By taking the Fourier transformation in $t$ and $\phi$ of the functions $c_i$%
, 
\begin{align*}
c_i(t,\phi,X,\theta) = \int \hat{c_i}(\omega,k,X,\theta) e^{-i(\omega t + k
\phi)} d\omega d k \ ,
\end{align*}
the equation becomes 
\begin{align*}
-\Delta \hat{c}_i+(k^2- V) \hat{c}_i =L^2 \omega^2 \hat{c}_i \ ,
\end{align*}
where the eigenvalue $k=m/(2\pi)$ is the wavenumber along the $\phi$%
-direction, with $m$ a non-vanishing integer, which follows from the range
of $\phi$.

According to Duhamel's principle \cite{Sternberg}, an inhomogeneous equation
for a function $W=W(x,t)$ of the form 
\begin{align*}
(d_t^2+M) W=f \ ,
\end{align*}
with $M$ a non-negative operator and initial conditions $W(\cdot,0)=\psi_1$, 
$\partial_t W(\cdot,0)=\psi_2$, has the following general solution 
\begin{align*}
W(\cdot,t)=\partial_t
B(t)\psi_1+B(t)\psi_2+\int_{0}^{t}B(t-\tau)f(\tau)d\tau \ , \quad B(t)=M^{-%
\frac{1}{2}}\sin(t M^{\frac{1}{2}}) \ .
\end{align*}
In our case, to ensure that the perturbed Maxwell equation can be solved we
need to demand that $V_{\text{eff}}>0$, with $V_{\text{eff}}=(\frac{m}{2\pi}%
)^2-V$. Since $V$ takes its maximum value when $\sin^2(\alpha)
\sin^2(q\theta) = 1$, we get to the following relation 
\begin{align*}
L < \frac{m}{4\pi} \frac{1}{\sqrt{e K}} \ .
\end{align*}
We can see that there is a maximum value of the size of the box $\sim 10 fm$.

\vspace{.5cm}

The complete stability analysis requires to study the most general
perturbations of the solutions of the present work, but this is a very hard
task even numerically as it involves a coupled system of linear PDEs. We
would like to revisit the perturbative stability of these solutions in a
future publication.

\subsection{Isospin perturbations}

Now, we will shortly analyze the \textquotedblleft Isospin perturbations'':
namely, the perturbations of the Isospin degrees of freedom. This analysis
will clarify why, in the present case, the \textquotedblleft most dangerous
perturbations'' are the ones of the profile in Eq. (\ref{pert}). Isospin
perturbations of the $SU(2)$ valued field $U$ defined in Eqs. %
\eqref{general1} and \eqref{good1} can be parametrized as (see \cite{14}): 
\begin{gather}
U \rightarrow U^{A}=A^{-1}UA\ ,\ A\in SU(2)\ ,  \label{isopert1} \\
A = A\left( \tau \right) \ ,\ \tau =\frac{t}{L}-\phi \ ,  \label{isopert2} \\
A^{\pm 1}(x^{\mu })=\cos \left( \varsigma _{0}\left( \tau \right) \right) 
\mathbf{1}_{2}\pm \sin \left( \varsigma _{0}\left( \tau \right) \right) 
\widetilde{n}^{i}t_{i}\ ,  \label{isopert3} \\
\widetilde{n}^{1}=\sin \varsigma _{1}\left( \tau \right) \cos \varsigma
_{2}\left( \tau \right) \ ,\ \ \widetilde{n}^{2}=\sin \varsigma _{1}\left(
\tau \right) \sin \varsigma _{2}\left( \tau \right) \ ,\ \ \widetilde{n}%
^{3}=\cos \varsigma _{1}\left( \tau \right) \ .  \label{isopert4}
\end{gather}%
In the present case, a small perturbation should be characterized by an $%
A\left( \tau \right) $ close to the identity. Thus 
\begin{gather*}
\varsigma _{0}\left( \tau \right) = \varepsilon \varsigma \left( \tau
\right) \ ,\ \ 0<\varepsilon \ll 1\ , \\
A^{\pm 1}(x^{\mu }) = \mathbf{1}_{2}\pm \varepsilon \varsigma \left( \tau
\right) \widetilde{n}^{i}t_{i}\ .
\end{gather*}%
Note that the only change that needs to be done with respect to classic
procedure of \cite{14} is the use of the light-cone time (instead of the
usual time-like coordinate in \cite{14}). The reason is that in \cite{14}
the authors considered the following type of Isospin perturbations%
\begin{eqnarray*}
U_{0} \rightarrow U_{0}^{A}=A^{-1}U_{0}A \ , \qquad A=A\left( t\right) \in
SU(2) \ ,
\end{eqnarray*}%
of the original static Skyrmions $U_{0}$ introduced by Skyrme himself:%
\begin{equation*}
\frac{\partial }{\partial t}U_{0}=\left( \frac{\partial x^{\mu }}{\partial t}%
\right) \frac{\partial }{\partial x^{\mu }}U_{0}=0\ .
\end{equation*}%
For the same reason, in the present case is more convenient to choose the
light-cone time.

A perturbation of the type defined in Eqs. (\ref{isopert1}), (\ref{isopert2}%
), (\ref{isopert3}) and (\ref{isopert4}) is called \textit{Isospin
perturbation} because of the following reason: 
\begin{gather*}
U(x^{\mu })=\cos \left( \alpha \right) \mathbf{1}_{2}+\sin \left( \alpha
\right) n^{i}t_{i} \  \\
\Rightarrow U^{A}=A^{-1}UA=\cos \left( \alpha \right) \mathbf{1}_{2}+\sin
\left( \alpha \right) \left( A^{-1}n^{i}t_{i}A\right) \ .
\end{gather*}%
Therefore, the perturbations of the $SU(2)$ valued field $U$ defined in Eqs. %
\eqref{general1} and (\ref{good1}) that we are considering in the present
subsection are fundamentally different from the perturbations in Eq. (\ref%
{pert}) since the latter only changes the profile $\alpha (X)$ while they
keep the functions $\Theta $ and $\Phi $ unchanged while the perturbations
in Eqs. (\ref{isopert1}), (\ref{isopert2}), (\ref{isopert3}) and (\ref%
{isopert4}) do not change the profile while they only affect the functions $%
\Theta $ and $\Phi $. Since we are studying linear perturbations, it makes
sense to analyze these two different types of perturbations separately.

It is direct but rather cumbersome to check that the perturbations on the
functions $\Theta$ and $\Phi$ that follows the criteria of \cite{14}, this
is, in the case with $t$-dependence as well when they depend on the
light-cone coordinate $\tau=t/L-\phi$, are in both cases zero-modes of the
system, meaning, the energy of the system does not decrease.

\end{document}